\documentclass[Times,twocolumn,tighten,deluxetable]{aastex63}
\makeatletter
\let\frontmatter@title@above=\relax
\usepackage{graphicx}
\usepackage{multirow}
\usepackage{txfonts}
\usepackage{natbib}
\usepackage{url}
\usepackage{hyperref}
\usepackage{longtable}
\usepackage{tabularx}
\usepackage{lipsum}
\usepackage{array}
\usepackage{rotating}
\usepackage{changepage}

\shorttitle{X-ray Properties of Swift J1727.8-1613}
\shortauthors{K. Chatterjee et al.}

\begin{document}

\title{{\it Insight}-HXMT View of the BHC Swift J1727.8-1613 during its outburst in 2023}

\correspondingauthor{Kaushik Chatterjee}
\email{kaushik@ynu.edu.cn, mails.kc.physics@gmail.com}

\author[0000-0002-6252-3750]{Kaushik Chatterjee}
\affiliation{South-Western Institue For Astronomy Research, Yunnan University, University Town, Chenggong, Kunming 650500, China}

\author[0000-0003-0793-6066]{Santanu Mondal}\email{santanu.mondal@iiap.res.in}
\affiliation{Indian Institute of Astrophysics, II Block Koramangala, Bengaluru 560034, Karnataka, India}

\author[0000-0002-7782-5719]{Chandra B. Singh}\email{chandrasingh@ynu.edu.cn}
\affiliation{South-Western Institue For Astronomy Research, Yunnan University, University Town, Chenggong, Kunming 650500, China}

\author[0000-0002-1190-0720]{Mutsumi Sugizaki}
\affiliation{National Astronomical Observatories, Chinese Academy of Sciences, 20A Datun Road, Chaoyang District, Beijing 100012, China}
\affiliation{Kanazawa University, Kakumamachi, Kanazawa, Ishikawa 9201192, Japan}


\begin{abstract}

The transient Galactic black hole candidate Swift\,J1727.8-1613 went through an outburst for the very first time in August 2023 and lasted for almost 6 months.
We study the timing and spectral properties of this source using publicly available archival {\it Insight}-HXMT data for the first 10 observation IDs that last from MJD 
60181 to 60198 with a total of 92 exposures for all three energy bands. We have detected quasi-periodic oscillations (QPOs) in a frequency range of $0.21 
\pm 0.01$ - $1.86 \pm 0.01$ Hz by fitting the power density spectrum. Based on the model-fitted parameters and properties of the QPOs, we classify them as type-C in 
nature. We also conclude that the origin of the QPOs could be the shock instabilities in the transonic advective accretion flows around black holes. The spectral analysis 
was performed using simultaneous data from the three on-board instruments LE, ME, and HE of \textit{Insight}-HXMT in the broad energy band of $2-150 $ keV. To achieve the 
best fit, spectral fitting required a combination of models e.g. interstellar absorption, power-law, multi-color disk-blackbody continuum, Gaussian emission/absorption, and 
reflection by neutral material. From the spectral properties, we found that the source was in an intermediate state at the start of the analysis period and was transitioning 
to the softer states. The inner edge of the accretion disk moved inward in progressive days following the spectral nature. We found that the source has a high inclination of 
$78^\circ-86^\circ$. The hydrogen column density from the model fitting varied in the range of $(0.12 \pm 0.02 - 0.39 \pm 0.08)\times10^{22}$ cm$^{-2}$.

\end{abstract}


\section{Introduction}

Stellar-mass black holes (SBHs) are one of the end products of the death of massive stars. They reside in binary systems where a companion star supplies matter and during accretion 
they emit radiation.  A significant amount of their radiation comes out in the form of X-rays, therefore, they are called black hole X-ray binaries (BHXRBs). Depending on the mass 
of the companion star, they can be mainly of two types: low-mass X-ray binaries (LMXRBs) or high-mass X-ray binaries (HMXRBs). The LMXRBs consist of type-A or later type of stars, 
whereas the HMXRBs consist of giant O or B-type of stars (e.g., White et al. 1995; Tetarenko et al. 2016). During the transfer of mass from the companion, matter accumulates at a 
distance from the BH due to insufficient viscosity to drive the matter inward (Mineshige 1993; Kuulkers et al. 1997; Parmar et al. 1997; Lasota 2001), which is called the pile-up 
radius (Chakrabarti et al. 2019; Chatterjee et al. 2022). Due to the accumulation of more matter, the temperature of the disk rises. The gradual increase in temperature 
develops instability at the pile-up radius, which further increases the viscosity. At some point in time when the accumulated matter gains enough viscosity to push forward, accretion 
starts. On the other hand, irradiation of the accumulated matter at the outer boundary by the central object can also trigger mass accretion when the temperature of the accumulated 
matter crosses a critical limit (King \& Ritter 1998; Mondal 2020).  Thus, these systems make ideal laboratories for studying the physics of accretion. BHXRBs mostly stay in a quiet 
dormant state when their flux remains at a very low detection level. Episodically, the radiation level increases by a very large factor and they become easily detectable, which is 
known as an outburst (Tanaka \& Shibazaki 1996; Remillard \& McClintock 2006). The X-ray luminosity increases by several orders of magnitude during an outburst when compared to a 
quiescence.  Depending on the nature of the outburst, the sources are classified into two types: persistent and transient sources. Most of the BHXRBs are transient type of sources. 
They stay in the dormant state for most of their lifetimes. The luminosity becomes $\sim10^{37-38} $ erg s$^{-1}$ during episodic outbursting phase (Tanaka \& Shibazaki 1996).

During the onset of an outburst, the flux in the light curve changes noticeably. Based on its nature, the outbursts are divided into two types (Debnath et al. 2010): 
fast rise slow decay (FRSD) and slow rise slow decay (SRSD). However, there are also some other classified outburst types, e.g., according to Zhang et al. (2019), outbursts can 
also be classified into glitch, reﬂare, multipeak, and mini-outburst. There are generally four designated spectral states during a complete BH outburst (Remillard \& McClintock 2006). 
They are known as the hard state (HS), hard intermediate state (HIMS), soft intermediate state (SIMS), and soft state (SS). Generally, when an outburst starts, it starts in the HS.
Then it slowly moves toward the SS through the HIMS and SIMS (Belloni et al. 2005, 2011, and references within). After it reaches its SS, the rising phase generally comes to 
an end and the outburst starts its decaying phase. In the decaying phase, it goes in the opposite direction and reaches the declining HS. When a BH outburst goes through all these 
states, it is known as a type-I outburst. When the soft state is absent, it is known as a `failed' or type-II outburst (Hynes et al. 2000; Brocksopp et al. 2001; Belloni et al. 2002; 
Brocksopp, Bandyopadhyay \& Fender 2004; Capitanio et al. 2009; Curran \& Chaty 2013; Del Santo et al. 2016; Tetarenko et al. 2016; Garcia et al. 2019; Alabarta et al. 2021). 

The energy spectrum of a BHXRB generally consists of two main components. One is the soft thermal multi-colour blackbody component, which is due to the blackbody radiation 
of the seed photons in the standard accretion disk (Shakura \& Sunyaev 1973; Novikov \& Thorne 1973). The other is the hard power-law component, which can extend up to very high 
energies. The origin of this component is thought to be due to the inverse Comptonization of a fraction of seed photons from the disk, which is intercepted by a hot Compton 
cloud (Sunyaev \& Titarchuk 1980, 1985; Haardt \& Maraschi 1993; Zdziarski et al. 1993; Titarchuk 1994; Chakrabarti \& Titarchuk 1995; Życki et al. 1999). When the outburst is in 
the HS, there is a very small contribution from the thermal blackbody component and mostly the (non)thermal powerlaw component dominates. In the SS, the thermal 
blackbody component becomes dominant while in the intermediate states (HIMS and SIMS), the contribution from the two components stays comparable. Other than these two components, 
in the case of a high soft state (HSS), there could be the presence of bulk motion Comptonization (BMC) which takes place due to the relativistic speed of matter when it reaches a 
closer to the BH (Blandford \& Payne 1981a,b; Payne \& Blandford 1981; Chakrabarti \& Titarchuk 1995; Psaltis \& Lamb 1997; Borozdin et al. 1999; Psaltis 2001). It was recently 
observed by Chatterjee et al. (2023) that BMC was present in the radiation spectrum of the black hole candidate (BHC) MAXI J0637-430. 

In addition, reprocessing of the Comptonized photons from the Compton cloud by the accretion disk is known as reprocessed radiation (George \& Fabian 1991; Ross \& Fabian 2005;
Garcia \& Kallman 2010), which generates reflection spectra. As a result, the Fe $K\alpha$ emission line at $\sim 6.4 $ keV and a reflection hump above 20 keV may originate. 
The presence of these complex features in the spectrum of BHXRBs makes it interesting to study. It is important to study and model these features to understand their origin. 
This can be achieved by fitting broadband data using multicomponent models. In this regard, high-resolution broadband data of {\it Insight}-HXMT (Zhang et al. 2020) can be very 
useful. 

Besides rich spectral features, temporal properties are equally important to understanding the dynamics of the accreting gas around the BHs. During an outburst, it has been observed 
that the light curves show changes in very small timescale, especially in the high-energy bands. The Fourier transformation of the light curve imprints such changes 
as both broadband noise and narrow features in the power density spectrum or PDS (van der Klis 1989). The broadband noise is spread over a large frequency range modeled by power-law 
or broad Lorentzian functions. The peak-like feature is a power peak in narrow frequency ranges, known as the quasi-periodic oscillation (QPO). The PDS is modeled by one or multiple 
Lorentzian profiles along with the power-law model. The Lorentzian model helps estimate the properties of the QPO (Nowak 2000; Belloni et al. 2002). Low frequency quasi 
periodic oscillations (LFQPOs) lie in the frequency range 
of a few mHz to $\sim 30$~Hz (Belloni et al. 2002; Casella et al. 2005; Motta et al. 2015). Depending on their properties, e.g., the frequency ($\nu$), $Q$-value ($=\nu/\delta\nu$, 
where $\delta\nu$ is the full width at half maximum or FWHM), (\%)RMS, etc., LFQPOs are classified into three different types: type A, B, and C (Casella et al. 2005). Several ideas 
were put forward to explain the LFQPO phenomenon in BHs. Although type-C LFQPO is now a well-studied phenomenon, there is still debate about its origin, as several models can explain 
it. Some of the often used explanations are the Lense-Thirring precession (Stella et al. 1999; Ingram et al. 2009), magneto-acoustic waves (Titarchuk et al. 1998), accretion-ejection 
instability (Tagger \& Pellat 1999) and the shock oscillation model (Molteni et al. 1996; Chakrabarti et al. 2005, 2008, 2015). However, compared to type-C, the other two types' origin 
is still not clearly known. 

The shock oscillation is one of the features of the two-component advective flow (TCAF; Chakrabarti \& Titarchuk 1995) model. This model suggests that there are two different forms of 
angular momentum distribution in matter supplied from the companion star, Keplerian and sub-Keplerian. Due to its high viscosity, the Keplerian component moves in a viscous timescale 
and forms a thin accretion disk on the equatorial plane. The accretion disk moves inwards as this matter gains critical viscosity. Less viscous accretion flow characterizes the sub-Keplerian 
one, which falls radially in a free fall timescale. This matter resides above and below the Keplerian disk. The sub-Keplerian component, which is optically thin, creates a shock front 
where infalling matter virtually stops because of the tug of war between centrifugal and gravitational forces. The boundary layer, sometimes referred to as the CENBOL (centrifugal 
pressure-supported boundary layer, Chakrabarti \& Titarchuk 1995), is the shock front. In this model, the post-shock region is the `Compton' cloud region, the repository of hot 
electrons. As blackbody photons are produced by the disk, the soft multicolor blackbody component can be explained by the Keplerian component. Some of the captured soft disk photons are 
upscattered and released as hard power-law photons by the CENBOL, which serves as the Comptonizing 
region. During the start of an outburst, the CENBOL forms farther away from the BH, and the disk is also truncated at a very large radius. As a result, photons originating from the disk become hard 
through inverse Compton scattering by hot electrons, and thus we always observe a hard state during the onset of an outburst from BHs. As time progresses, the Keplerian disk moves inwards, producing 
more blackbody radiation. More soft photons are intercepted in the CENBOL, which are upscattered and help to cool down the CENBOL. More photons are emitted in the whole process and flux increases 
and the source goes toward a softer state. In a high soft state, the CENBOL is totally quenched, and the disk reaches very close to the black hole. Thus, only the disk photons now contribute to the 
radiation, which makes the spectrum soft.

Besides describing the spectral features and their changes, this model also explains the QPO properties. In the HS, when the CENBOL is bigger, it could produce LFQPOs due to the 
resonance oscillation of the shock, when the infall and cooling timescales match (Molteni et al. 1996, Chakrabarti et al. 2015, Garain et al. 2014). As the outburst progresses, CENBOL loses 
its size, and we see a rise in the frequency of the QPO. This is discussed with mathematical relation in the discussion section. As the SS approaches, the CENBOL shrinks in size, and there 
is no oscillation, cooling takes over the heating timescale. Thus, we do not see any QPO in the SS. In between the HS and SS, when the CENBOL is intermediate in size, there 
could be the presence of either type-C or type-A/B QPOs. In the HIMS, we see type-C, whereas, in the SIMS, mostly type-A/B is seen.

As changes in both the spectral and timing properties come from the same system and changes in spectral states can also dictate the types of QPOs, there could be a possible correlation 
between the spectral and temporal properties. A connection between these two properties is observed using a hardness intensity diagram, or HID (Homan et al. 2001), which shows the variation of 
the flux in the light curve (considered here as timing property) with the hardness ratio, or HR, which shows roughly the variation of the spectral nature of the outburst (considered as 
spectral property here). A detailed outburst study showed that the changes in both properties are due to the interplay between different mass accretion rates or variation in viscosity, which can 
be understood through the accretion rate ratio intensity diagram or ARRID (Mondal et al. 2014; Jana et al. 2016; Chatterjee et al. 2020). The RMS-intensity diagram (RID, Munoz-Darias et al. 2011), 
and hardness ratio-intensity diagram or HRD (Belloni et al. 2005) can also link spectral and temporal properties from a pure observational ground. As shown in Chakrabarti et al. (2015), 
when the cooling timescale is within 50\% of the infall/heating timescale, the QPO originates, which is in accord with the spectral model fitted ARRID in Mondal et al. (2014), where QPOs 
disappeared for ARR $<0.5$ in the ARRID. Unlike most of the explanations in the literature, the shock oscillation model in TCAF paradigm is successful in addressing the spectral and temporal properties 
responsible for the origin of QPOs, which prompted us to consider this model for rest of the paper to explain QPOs observed in Swift J1727.8-1613 during {\it Insight}-HXMT era.

The BHC Swift J1727.8-1613 was discovered very recently on 24 August 2023 or MJD 60180 (Kennea \& Swift Team 2023). MAXI (Matsuoka et al. 2009) detected  the source to be very bright with a 
flux of $7 $ Crab in the $2-20 $ keV energy range. Since then multi-wavelength follow-up observations were carried out by several ground-based and space-bourne telescope facilities (Baglio 
et al. 2023; Miller-Jones et al. 2023; Negoro et al. 2023b; O’Connor et al. 2023; Wang \& Bellm 2023; Williams-Baldwin et al. 2023). The source is reported to be located at a distance of 
$2.7 \pm 0.3$ kpc (MataSanchez et al. 2024) and a mass of $10 \pm 2 ~M_\odot$ (Svoboda et al. 2024). The revelation of the presence of hydrogen and helium emission lines in the optical 
spectrum led to its classification as an LMXRB (Castro-Tirado et al. 2023). The X-ray spectrum and other properties point to its nature as a BHXRB (Liu et al. 2023; Sunyaev et al. 2023). 
This was supported strongly after the detection of type C QPOs  (Palmer \& Parsotan 2023; Draghis et al. 2023; Bollemeijer et al. 2023; Mereminskiy et al. 2023) and a flat radio spectrum 
(Miller-Jones et al. 2023; Bright et al. 2023). There have been X-ray monitoring of the source with {\it NICER} (O'Connor et al. 2023), {\it NuSTAR} (Dovciak et al. 2023), {\it Insight}-HXMT
(Peng et al. 2024), {\it IXPE} (Veledina et al. 2023), {\it AstroSat} (Katoch et al. 2023). Using {\it IXPE} data on September 7, 2023, Veledina et al. (2023) estimated its polarization 
with a polarization degree (PD) of $4.1\% \pm 0.2\%$ and polarization angle (PA) of $2^\circ.2 \pm 1^\circ.3$. There was a further polarization study of the source by Ingram et al. (2023). 

At the begining of the outburst, the source was consistent with being in the hard state, described by a power-law with a photon index ($\Gamma$) $\sim 1.5-1.7$ (Liu et al. 2023). 
Later, using the {\it NICER} observations, Bollemeijer et al. (2023) reported the softening of the spectrum with a substantial contribution of the soft blackbody component after 25 days from 
the start of the outburst. The presence of LFQPOs was reported by Draghis et al. (2023) and the evolution of LFQPOs at the high energy band using {\it AstroSat} data was reported by Nandi 
et al. (2024). The source also showed soft time lag of $0.014 \pm 0.001$ sec between energy bands of $3-10$ and $0.5-3$ keV on August 29 2023 (MJD 60185) (Debnath et al. (2023) using {\it 
NICER} data). With reflection spectroscopy, these authors found the disk inclination to be $\sim 87^\circ$. However, none of the studies in the literature discussed the origin of the 
spectral and temporal properties in the context of the shock oscillation model, which prompted us to analyze the broadband X-ray data from {\it Insight}-HXMT and interpret them in light of 
accretion with shocks.

The paper is organized as follows: in \S2, we describe the observation, data selection, reduction, and analysis procedures. In \S3, we portray the spectral and temporal properties and observed 
results. In \S4, we discuss the possible physical origin of our result and connection between observed features. Finally, in \S5, we summarize the results and draw conclusions.

\section{Observation, Data Selection, Reduction, and Analysis}

The source is being extensively monitored by various X-ray satellites, as discussed earlier. For this study, we use China's first dedicated X-ray satellite {\it Insight}-HXMT (HXMT; Zhang et al. 
2014, 2020) data. In the subsequent subsections, we discuss the data selection, reduction, and analysis, respectively.

\subsection{Data Selection}

The source was regularly monitored by the {\it Insight}-HXMT satellite. In the \href{http://archive.hxmt.cn/proposal}{data archive}, observations are available on-demand. We found that there 
are a total of 34 observation IDs available \footnote{Although 3 of them were not downloadable}. Out of those available 31  observation IDs, we took the first 10 observation IDs. However, unlike 
other X-ray satellite data, each observation ID in {\it HXMT} has several exposures. Also, some of the observation IDs have continuous observations of about 2 to 3 days. We found that for these 
first 10 observation IDs, each of them has a minimum of 7 to a maximum of 18 exposures. Initially, we perform the analysis of all the exposures for the first 3 observation IDs and found no 
significant change in the parameters of spectral fits in the consecutive exposures. Therefore, we avoid fitting every exposure of each observation ID. The details of the selected data are given 
in Table 1.

Here, we note that, in general, the exposure is the time span of observation within an observation ID. The {\it Insight}-HXMT
has several observation IDs for this source, each of which lasts for a different amount of time, which is the exposure time. In Table 1, one can see that all the observation IDs have different 
exposures, with some of them having a time gap between them. When reducing the data, we found that inside every observation ID, there were several sub-IDs, the name of which starts with that 
particular ID. The total exposure of an entire observation ID was cut into several sub-IDs with different exposure times but with no time gap between them. This can be seen in Table 3 (columns 
3 and 5). GTIs are also different. They can be separated in time between the consecutive ones. Here, this is not the case. Thus, to differentiate between them, we called these  as `exposure 
IDs', and used this phrase throughout the paper. Within 1 observation ID (like the first one in Table 1), there are several exposure IDs (like 12 exposure IDs in the 1st column of Table 3). 
This has not been defined like this before. To keep things easy, we have taken this approach.

\begin{table}
\scriptsize
 \addtolength{\tabcolsep}{-1.5pt}
 \centering
 \caption{List of observations used for this work. Here, column 1 lists the observation IDs. Columns 2 and 3 denote the start and end date of observations, while columns 4 and 5 refer to corresponding 
        MJDs. Column 6 gives the exposure time of each observation.}
 \label{tab:table1}
 \begin{tabular}{cccccc}
 \hline
       Obs. Id.     &    Start Date       &       End Date       &   Start MJD       &     Stop MJD       &      Exp. (s)      \\
         (1)        &       (2)           &         (3)          &      (4)          &       (5)          &         (6)        \\
\hline 
     P0614338001    &    2023-08-25       &      2023-08-27      &    60181.34       &     60183.05       &       148495       \\
     P0614338002    &    2023-08-27       &      2023-08-28      &    60183.05       &     60184.23       &       137343       \\
     P0614338003    &    2023-08-29       &      2023-08-31      &    60185.30       &     60187.15       &       159958       \\
     P0614338004    &    2023-08-31       &      2023-09-02      &    60187.15       &     60189.07       &       165684       \\
     P0614338005    &    2023-09-02       &      2023-09-04      &    60189.07       &     60191.06       &       171384       \\
     P0614338006    &    2023-09-04       &      2023-09-06      &    60191.06       &     60193.44       &       206034       \\
     P0614338008    &    2023-09-07       &      2023-09-08      &    60194.03       &     60195.09       &        91378       \\
     P0614338009    &    2023-09-08       &      2023-09-09      &    60195.09       &     60196.08       &        85659       \\
     P0614338010    &    2023-09-09       &      2023-09-10      &    60196.08       &     60197.07       &        85653       \\
     P0614338011    &    2023-09-10       &      2023-09-11      &    60197.07       &     60198.06       &        85650       \\
\hline 
 \end{tabular}
\end{table}

\subsection{Data Reduction}

The main scientific instrument of {\it Insight}-HXMT is an array of 18 NaI/CSI phoswich scintillation detectors, each of which covers an effective area of 286 cm$^{2}$. The satellite has three 
instruments or payloads: High Energy (HE; Liu et al. 2020), Medium Energy (ME; Chen et al. 2020), and Low Energy (LE; Cao et al. 2020). They have an effective area of 5100, 952, and 384 cm$^{2}$ 
and cover an energy range of $20-250$, $5-30$, and $1-15 $ keV, respectively.  

After downloading on-demand level-1 data from the archive, we produced cleaned level-2 data for science analysis. The processing of cleaning raw data was done in the following way. We first 
installed the \href{http://hxmten.ihep.ac.cn/software.jhtml}{HXMTDAS}\footnote{http://hxmt.org/index.php/usersp/dataan} v2.05. Using this software, we run the {\fontfamily{pcr}\selectfont
hpipeline} command using proper input and output directories. This pipeline performs several subsequent automatic commands for all three instruments under some given conditions. Some conditions 
were set to get a good time interval, e.g., elevation angle $> 10^\circ$; geomagnetic cutoff rigidity $> 8$ GeV; pointing offset angle $< 0.04^\circ$; $> 600$ sec away from the South Atlantic 
Anomaly (SAA). All these commands altogether extract and clean the raw data and produce cleaned and analyzable science products\footnote{This is discussed in the 
\href{http://hxmten.ihep.ac.cn/SoftDoc/501.jhtml}{HXMT manual}\footnote{http://hxmten.ihep.ac.cn/SoftDoc/501.jhtml} in full details}. The specific commands {\fontfamily{pcr}\selectfont 
hespecgen}, {\fontfamily{pcr}\selectfont mespecgen}, and {\fontfamily{pcr}\selectfont lespecgen} produce the spectra for HE, ME, and LE instruments. Whereas, the {\fontfamily{pcr}\selectfont 
helcgen}, {\fontfamily{pcr}\selectfont melcgen}, and {\fontfamily{pcr}\selectfont lelcgen} tasks produce the light curve files for the three instruments. {\fontfamily{pcr}\selectfont 
herspgen}, {\fontfamily{pcr}\selectfont merspgen}, and {\fontfamily{pcr}\selectfont lerspgen} produce corresponding response files. The background subtraction for both the timing and spectral 
data was performed using the commands {\fontfamily{pcr}\selectfont hebkgmap}, {\fontfamily{pcr}\selectfont mebkgmap}, and {\fontfamily{pcr}\selectfont lebkgmap} for instruments HE, ME, and 
LE, respectively. For the $\chi^2$ fit-statistics in {\texttt XSPEC}, we grouped the spectrum using the {\fontfamily{pcr}\selectfont grppha} task of FTOOLS to a minimum of 30 counts per bin. 
We also set the time bin size to $0.01$ sec to produce the light curves for generating PDS and QPOs. 

Using the cleaned spectra and light curves, we further fitted and analyzed to extract the outburst properties, which are discussed in the next section.

\subsection{Data Analysis}

We have downloaded the daily average light curve of \href{http://maxi.riken.jp/star\_data/J1727-162/J1727-162.html}{{\it MAXI/GSC}}\footnote{http://maxi.riken.jp/star\_data/J1727-162/J1727-162.html} 
from the archive to study the outburst profile in addition to {\it Insight}-HXMT. Using the unbinned light curves from LE, ME, and HE modules, we have first produced $0.01$ sec time-binned light curves. 
Then the power density spectrum (PDS) was generated using the fast Fourier transformation (FFT) technique in the {\fontfamily{pcr}\selectfont powspec} task of the {\fontfamily{pcr}\selectfont XRONOS} 
package in the {\fontfamily{pcr}\selectfont HEASoft} software. Each observation's data was subdivided into several intervals, each of which contains 8192 newbins. First, the PDS for each interval is 
generated and then they are averaged to make a resultant PDS. The PDS is normalized in a way that their integral can give the RMS-squared fractional variability. The white noise level is 
subtracted by using a normalization value of $-2$ in {\fontfamily{pcr}\selectfont powspec} and a geometrical rebinning of -1.02 is used. For some data, we needed to use a geometrical rebinning of 
-1.05 to have a prominent QPO-like nature in the PDS. We model the PDS with Lorentzian model for both the fundamental QPO and the harmonic components in {\fontfamily{pcr}\selectfont powspec}. The 
parameters from the fitted Lorentzian model for the fundamental QPO help us obtain the frequency ($\nu_{qpo}$), the full-width at half maximum or FWHM, and the normalization. For several 
observations, we have found the presence of single or multiple harmonics. We have also extracted their properties using the same fitting method. We have fitted the light curves of all the exposures 
of the listed 10 observations. We report them in the result section.

For the spectral analysis, we have used the same LE, ME, and HE modules and fitted the broadband data in the $2-150 $ keV energy band. The best is achieved by using the combination of 
{\fontfamily{pcr}\selectfont disk blackbody}, {\fontfamily{pcr}\selectfont power-law}, {\fontfamily{pcr}\selectfont gaussian}, and {\fontfamily{pcr}\selectfont pexrav} models. For some observations, 
we needed to add a {\fontfamily{pcr}\selectfont gabs} model to achieve the best fit. For interstellar absorption, we have used the {\fontfamily{pcr}\selectfont tbabs} model. Since we simultaneously 
fit all three modules, we have included a {\fontfamily{pcr}\selectfont constant} to normalize the three resultant fittings. The following two model combinations are used: 
i) {\fontfamily{pcr}\selectfont constant*tbabs*gabs*(diskbb + power-law + gaussian + pexrav)}, ii) {\fontfamily{pcr}\selectfont constant*tbabs*(diskbb + power-law + gaussian + pexrav)} (for which 
{\fontfamily{pcr}\selectfont gabs} was not required. The spectral analysis was done for the exposures are marked with `*' in Table 3. We did not perform the spectral analysis for all epochs 
as we did not see change in spectral features over such a small time gap unlike the timing properties, which showed variations within a day. We report them in the result section. 

We note that light curves for the LE instrument were not produced for some exposures. Therefore, we represented only those exposures' MJDs in Table 4 for which all three light curves were available 
for analysis for uniformity.

\section{Results}

Using the timing and spectral analysis presented above, we have studied the accretion flow properties of the very recent 2023 outburst of the BHC Swift J1727.8-1613. We studied the QPO properties 
of the source during the outburst as well as the spectral nature and radiation properties using {\it Insight}-HXMT data. We broadly discuss the results in the next two sections.

\subsection{Temporal Properties}

\subsubsection{Outburst Profile and Hardness Ratio}

\begin{figure}[!h]
\vskip 0.2cm
  \centering
    \includegraphics[width=8.5cm]{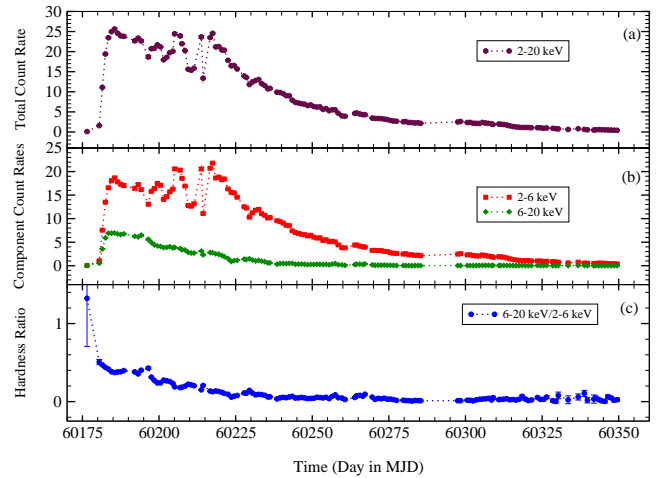}
    \caption{Variation of the (a) MAXI/GSC $2-20 $ keV count rate, (b) MAXI/GSC $2-6$ and $6-20 $ keV count rates, and (c) hardness ratio with time. The HR is the ratio of the $6-20 $ keV count 
	     rate to the $2-6 $ keV count rate of the MAXI/GSC data.}
\end{figure}

Figure 1 shows the variation of the flux obtained from {\it MAXI/GSC} during the outburst. The outburst started roughly at around MJD 60180 (2023 August 24), when its flux came out of quiescence. 
The {\it MAXI/GSC} flux started to rise after this date, as we can notice from panel (a) of Figure 1. Within five days, the flux increased very rapidly and the total flux reached its peak on MJD 
60185 (2023 August 29). After that, the flux started to decrease very slowly, except for the period from MJD 60200 to 60225, where the flux showed a significant change. In panel (b), we show the 
variations of the count rates with time in the $2-6$ and $6-20$ keV energy bands. We notice that, after MJD 60185, the hard flux (i.e., $6-20$ keV) decreased very slowly, whereas, the soft $2-6$ 
keV flux showed a lot of variation just like the total $2-20$ keV flux. On MJD 60185 the $2-20$ keV and hard $6-20$ keV fluxes reached their peak values, whereas, the soft flux reached its peak 
much later at around MJD 60217. This suggests that the variation in the total count rate is mainly due to the variation of the soft component. After this period, the soft flux also decreased slowly 
till the end of the outburst and entered again into the quiescence phase.

In panel (c), we show the variation of the hardness ratio, which is the ratio of the hard ($F_H$) $6-20 $ keV to the soft ($F_S$) $2-6 $ keV flux. At the start of the outburst, the HR was $\sim 
0.5$ and then it gradually decreased till MJD 60235 as the outburst progressed. After that, the HR became almost constant, although decreasing very slowly to $\sim 0.025$ at the end of the considered 
data of the outburst on MJD 60350. The way HR is defined, when this value is on the higher side, the spectral nature of the source is hard. This is the general case when an outburst is starting. 
When this value is low, there is dominance of soft flux over hard flux and the spectral nature would be soft. Without studying the detailed spectral analysis, HR provides a quick idea about the 
spectral nature of an outburst. Depending on the HR value, we propose that the source was already in the intermediate state after the onset of the outburst. After MJD 60185, it made a 
transition into the SS.

\subsubsection{Quasi Periodic Oscillations (QPOs)}

\begin{figure}[!h]
  \centering
    \includegraphics[width=6.0cm,angle=270]{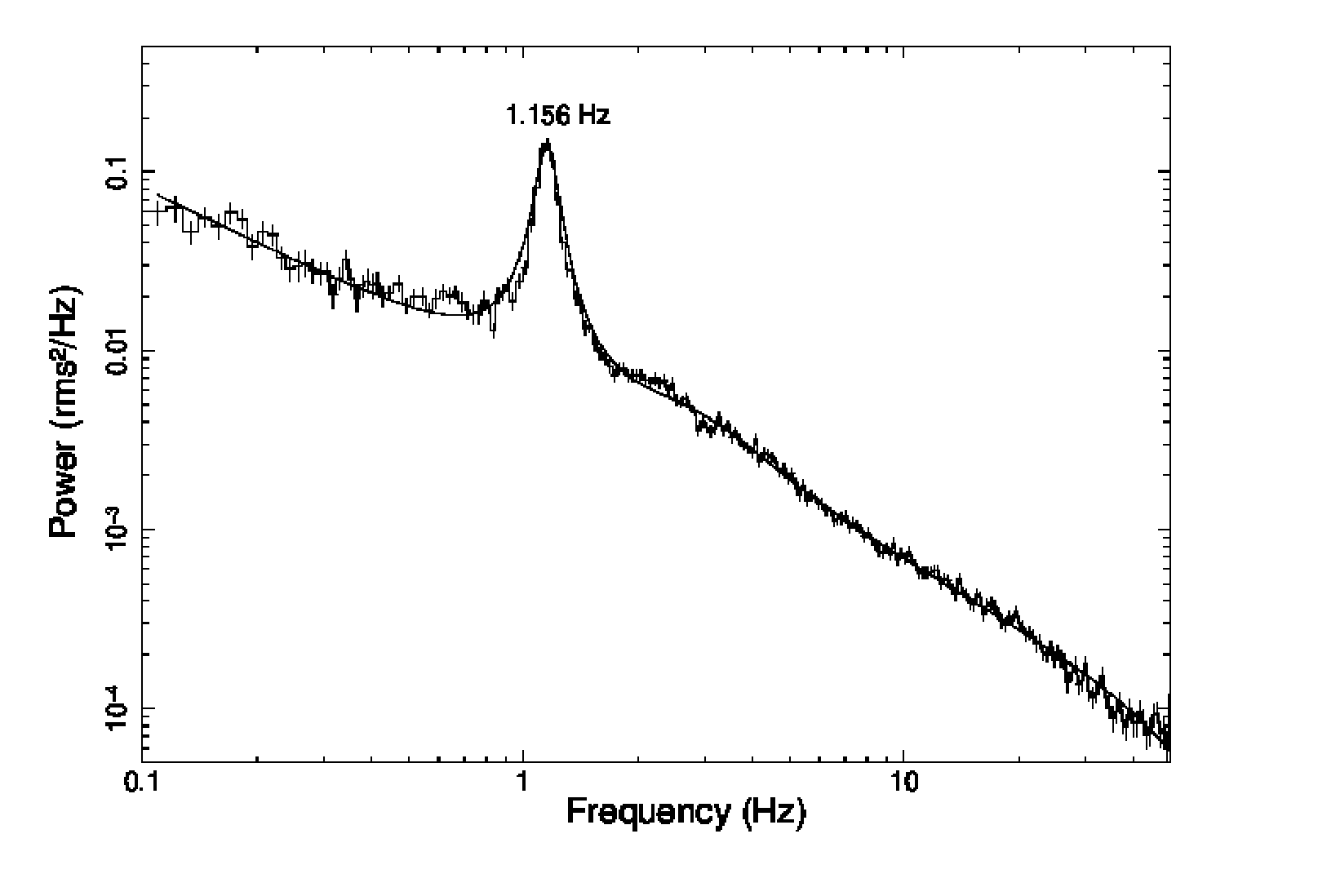}
    \caption{Model fitted PDS using $0.01$~sec time-binned HE light curve from the observation ID P0614338004 (Exposure ID P061433800401-20230831-01-01).}
\end{figure}

\begin{figure}[!h]
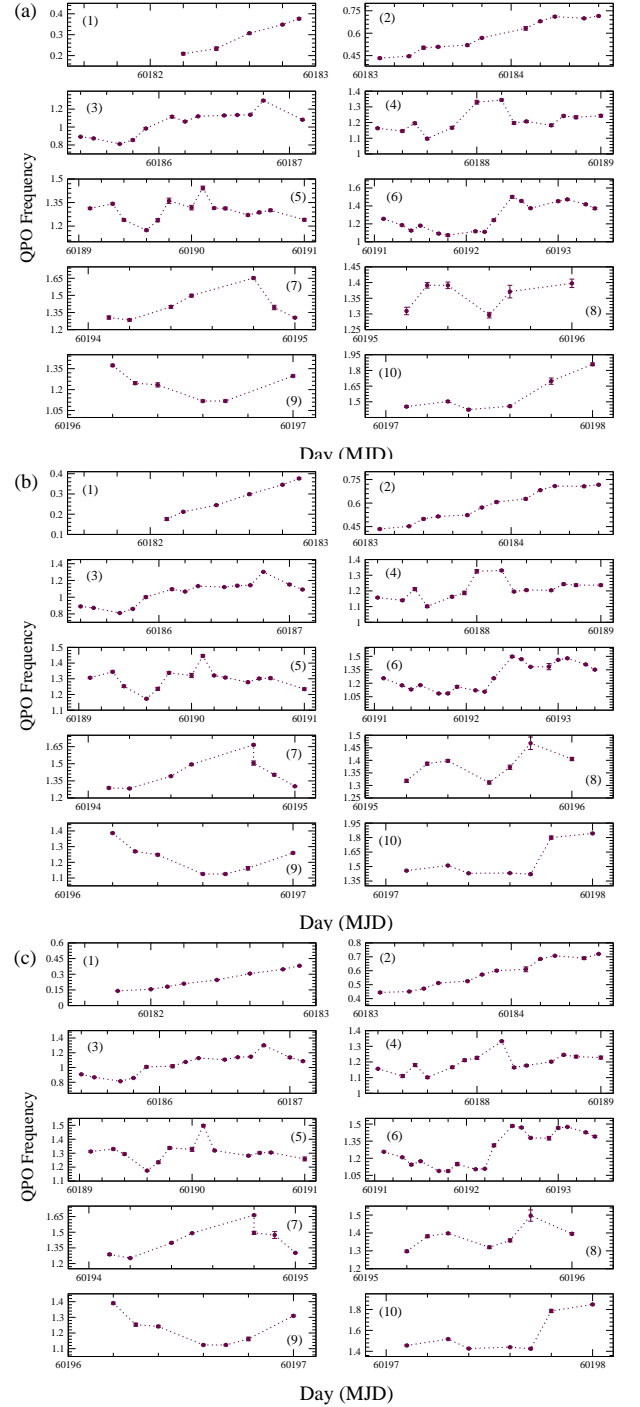

\centering
\vbox{
\includegraphics[width=8.5truecm,angle=0]{QPO_Exp_le.eps}\vskip 0.1cm
\includegraphics[width=8.5truecm,angle=0]{QPO_Exp_me.eps}\vskip 0.1cm
\includegraphics[width=8.5truecm,angle=0]{QPO_Exp_he.eps}
}
\caption{Evolution of QPO frequencies (Hz) with time. Here we have shown three different plots for (a) LE, (b) ME, and (c) HE instruments. Panels 1--10 in each of the three of these plots (a, b, c) 
represent the $\nu_{qpo}$ variation for the 10 observation IDs we used.}
\end{figure}

As described in \S2, we used the $0.01$ sec time-binned light curves from all three bands (LE, ME, and HE) and produced the PDS to study QPOs. We find that QPO nature was present in all the light 
curves in all three energy bands. In Figure 2, we show a model fitted PDS continuum in the $0.1-50$~Hz frequency range for $0.01$~sec time-binned HE light curve from the observation 
ID P0614338004 (Exposure ID P061433800401-20230831-01-01). In Figure 3, we show the variation of the QPO frequency ($\nu_{qpo}$) for the 10 different obs. IDs as stated before. We first checked 
the first obs. ID {\fontfamily{pcr}\selectfont P0614338001}. It has several exposure IDs (given in Table 3). While we fit every exposure of this obs ID, we found that the QPO frequency was changing even 
within this small period. This is the reason we fitted all the light curves of the exposures of the obs IDs that we have taken for analysis. We find that for every single obs ID, the $\nu_{qpo}$ showed 
significant variation. Figure 3(a-c) shows the variations of the $\nu_{qpo}$ with time for 10 different obs IDs for LE, ME, and HE respectively. It is interesting to observe that $\nu_{qpo}$ has changed 
in frequency so much within this short period, which suggests that there are some changes happening in the system in a very short timescale.

\begin{figure}[!h]
  \centering
    \includegraphics[width=8.0cm]{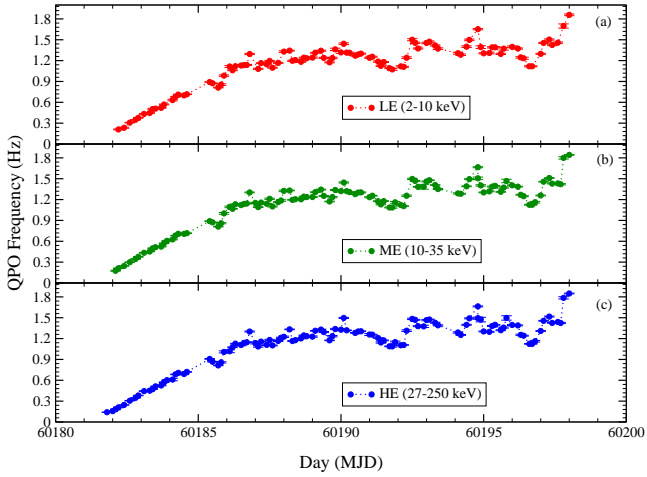}
    \caption{Evolution of QPO frequency with time during the whole period of analysis for (a) LE, (b) ME, and (c) HE.}
\end{figure}

In Figure 4, we show the evolution of $\nu_{qpo}$ for the total duration of our analysis period. In (a), (b), and (c) panels the variations are for LE, ME, and HE respectively. We find that the QPO
frequencies were $0.23 \pm 0.01$, $0.24 \pm 0.01$, and $0.24 \pm 0.01$ Hz for LE, ME, and HE on the starting day of our analysis. After this day, $\nu_{qpo}$ shows a rapid increase 
with time till $\sim$ MJD 60187 when $\nu_{qpo}$ was $1.16 \pm 0.01$, $1.16 \pm 0.01$, and $1.16 \pm 0.01$ for LE, ME, and HE, respectively. After this date, it increased very slowly 
till MJD $\sim 60190$ when $\nu_{qpo}$ was $1.44 \pm 0.01$, $1.44 \pm 0.01$, and $1.50 \pm 0.01$ for the three bands. Then it decreased for a very short period until it rapidly increased 
to $\sim 1.5$ Hz again in all three bands. After that, it increased and decreased in a narrow range of $\sim 1.0-1.5$ Hz till MJD $\sim 60197$, and then it finally increased again. We have 
discussed the possible reasons behind this nature in the next section. 

Using the PDS fitting, we extracted the centroid frequency, the full-width at half maximum (FWHM) and the Normalization (LN)) of the QPOs. The `{\fontfamily{pcr}\selectfont fplot}' task 
of the `{\fontfamily{pcr}\selectfont ftools}' package in the {\fontfamily{pcr}\selectfont HEASoft} software produces plots of the light curves. Using this task, we plotted and saved each value of time and 
the corresponding count rate. Then we average the counts to find out the average count rate for that light curve. The same task is applied to extract both the source and background count rates for all the 
exposures. Then, using these estimates, we calculated the $Q$-value and RMS that represent the sharpness of the QPO and the fractional variability in the PDS, respectively. These values are listed in Table 
4 in columns 5--7 ($Q$-value) and 8--10 (RMS) for LE, ME, and HE respectively.

\begin{figure}[h!]
\centering
\includegraphics[width=8.5cm]{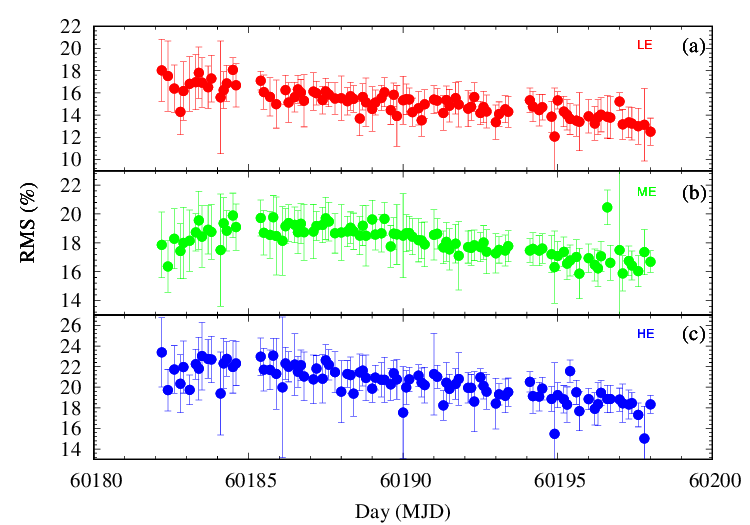}
\caption{Evolution of the QPO RMS (\%) with time (MJD) for (a) LE, (b) ME, and (c) HE.}
\end{figure}

In Figure 5, we show the evolution of the QPO RMS (\%) with time (MJD). In the LE band (Fig. 5a), the RMS was higher at the start of the analysis period and it then gradually 
decreased. In the ME band (Fig. 5b), it increased from the onset of the analysis period for some time and then decreased. The HE band (Fig. 5c) shows a similar trend as in LE band.
Since the QPO nature has already been reported in the recent paper by Yu et al. (2024), we are not focusing on the classification of the QPO. However, by the values 
of both $Q$-factor and RMS, it can be said that the QPO is type-C in nature for all the exposures reported here.

\subsection{Spectral Properties}

\begin{figure}[!h]
\centering
\vbox{
\includegraphics[width=5.0truecm,angle=270]{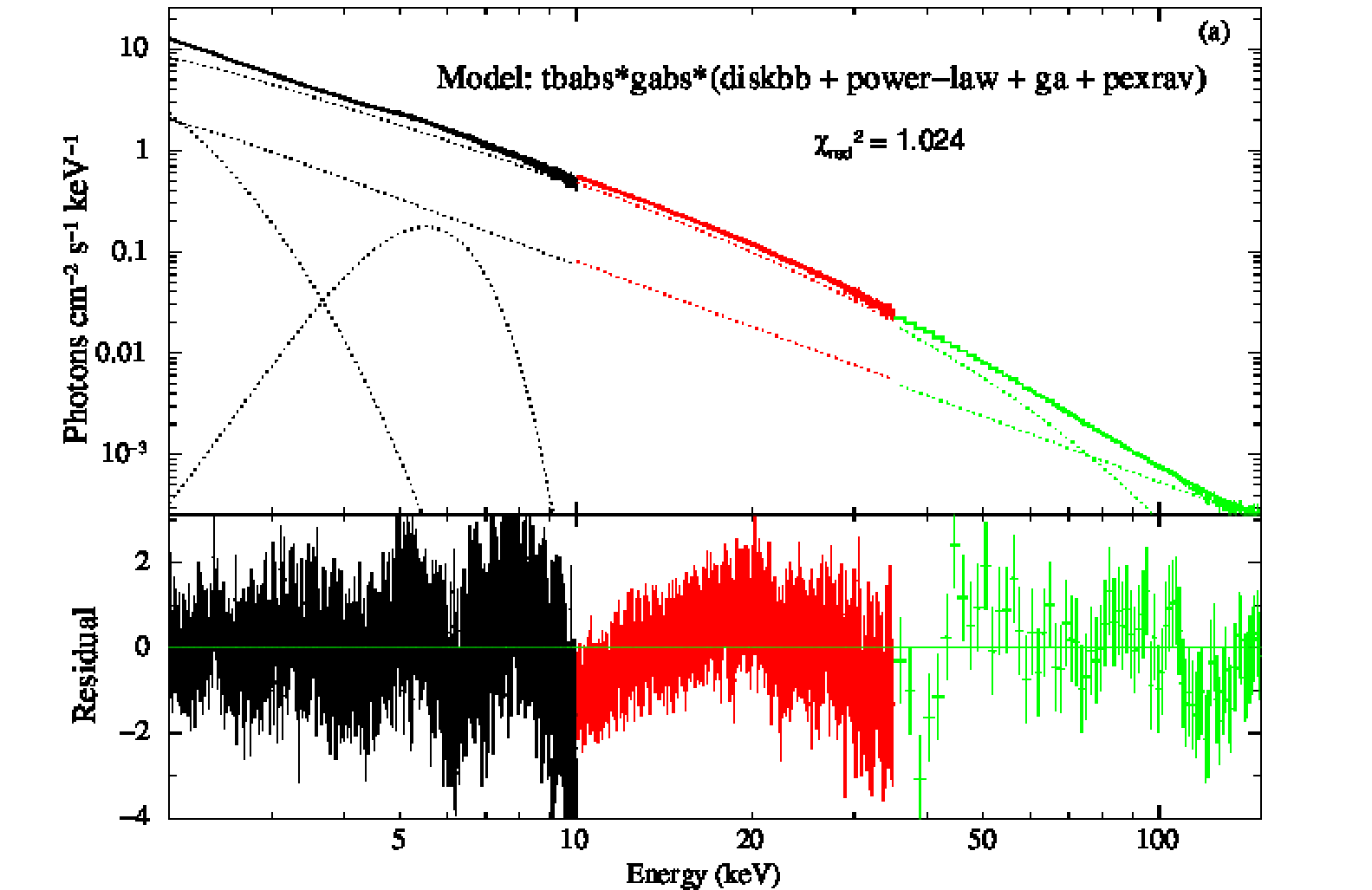}\hskip 0.5cm
\includegraphics[width=5.0truecm,angle=270]{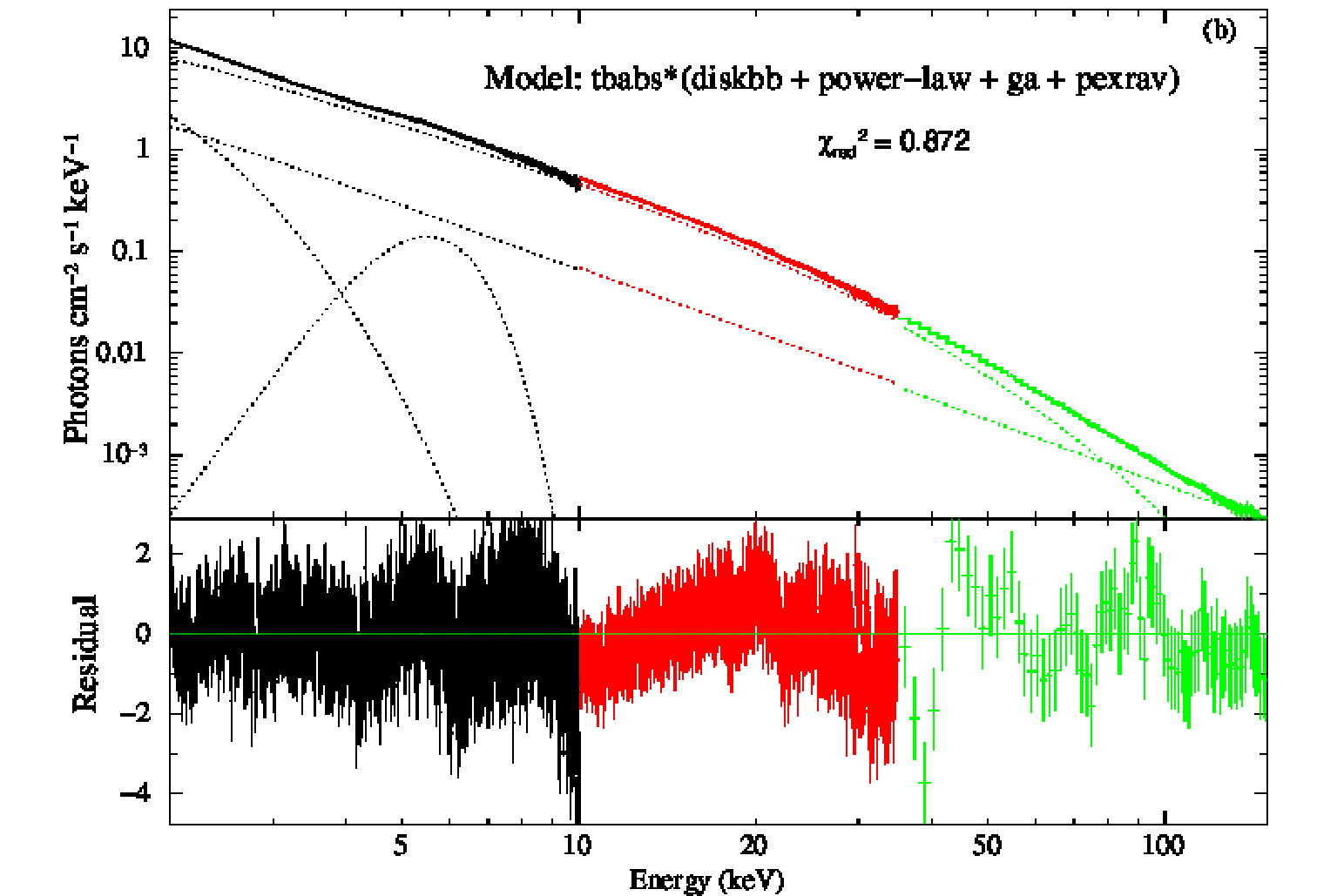}}
\caption{Model fitted spectrally analyzed unfolded spectra. (a) is or observation ID. P061433800501 (Exposure: P061433800501-20230902-01-01) for which the \textit{gabs} model was required to 
         achieve the best fit. (b) is for the observation ID. P061433800601 (exposure: P061433800601-20230904-01-01) for which the \textit{gabs} model was not required to achieve the best fit.}
\end{figure}

Along with the timing properties, studying the spectral properties provides more insights on the nature of the outburst. With the available {\it Insight}-HXMT data, we analyzed the source 
for a total of 32 exposures. These exposure IDs are marked with a `*' sign in the 1st column of Table 3. An extensive spectral analysis on this source has not been done so far. We perform 
a detailed spectral analysis on this source for the first time using HXMT data. We started our spectral analysis from MJD 60181.4. We have simultaneously used {\fontfamily{pcr}\selectfont 
LE+ME+HE} in the $2-150$ keV energy band (LE in 2--10, ME in 10--35, and HE in 35--150 keV) for our spectral fitting for all the selected exposure IDs.

We first fitted the data with the phenomenological \href{https://heasarc.gsfc.nasa.gov/xanadu/xspec/manual/node166.html}{{\fontfamily{pcr}\selectfont diskbb}} and 
\href{https://heasarc.gsfc.nasa.gov/xanadu/xspec/manual/node221.html}{{\fontfamily{pcr}\selectfont power-law}} models. We used the multiplicative 
\href{https://heasarc.gsfc.nasa.gov/xanadu/xspec/manual/node273.html}{{\fontfamily{pcr}\selectfont tbabs}} model to account for the interstellar absorption. Our model read as {\fontfamily{pcr}\selectfont
constant*tbabs(diskbb + powerlaw)} in {\fontfamily{pcr}\selectfont XSPEC}. The {\fontfamily{pcr}\selectfont constant} is taken to normalize between three different instruments in LE, ME, and HE. 
We found that the fit was not statistically acceptable. There was a clear sign of the Gaussian feature in the unfolded spectrum around $5-6$ keV. Also, there was a hump-like feature $\sim 20$ keV. 
Besides these features, the absorption feature was noticed below 2 keV. Thus, we refitted the data using the model combination as {\fontfamily{pcr}\selectfont constant*tbabs*gabs(diskbb + power-law + 
ga + pexrav)}. Here, the \href{https://heasarc.gsfc.nasa.gov/xanadu/xspec/manual/node251.html}{{\fontfamily{pcr}\selectfont gabs}} model takes care of the low energy absorption feature, while the 
\href{https://heasarc.gsfc.nasa.gov/xanadu/xspec/manual/node181.html}{{\fontfamily{pcr}\selectfont gaussian}} and \href{https://heasarc.gsfc.nasa.gov/xanadu/xspec/manual/node214.html}{{\fontfamily{pcr}\selectfont 
pexrav}} models take care of the contribution from the Gaussian and hump-like nature, as mentioned above. In Fig. 5a, we show the best-fitted unfolded spectrum using this model combination for the 
observation ID P061433800501 (Exposure ID: P061433800501-20230902-01-01), where $\chi^2/DOF \sim 1$. After some exposures, we found that the {\fontfamily{pcr}\selectfont gabs} model was not needed 
anymore to achieve the best fit. Then, the model combination in {\fontfamily{pcr}\selectfont XSPEC} reads as {\fontfamily{pcr}\selectfont constant*tbabs(diskbb + power-law + ga + pexrav)}. In Figure 
6b, we show an unfolded best-fitted spectrum using the later model combination. This is for the observation ID. P061433800601 (exposure ID: P061433800601-20230904-01-01), for which $\chi^2/DOF \sim 0.9$.

\begin{figure}
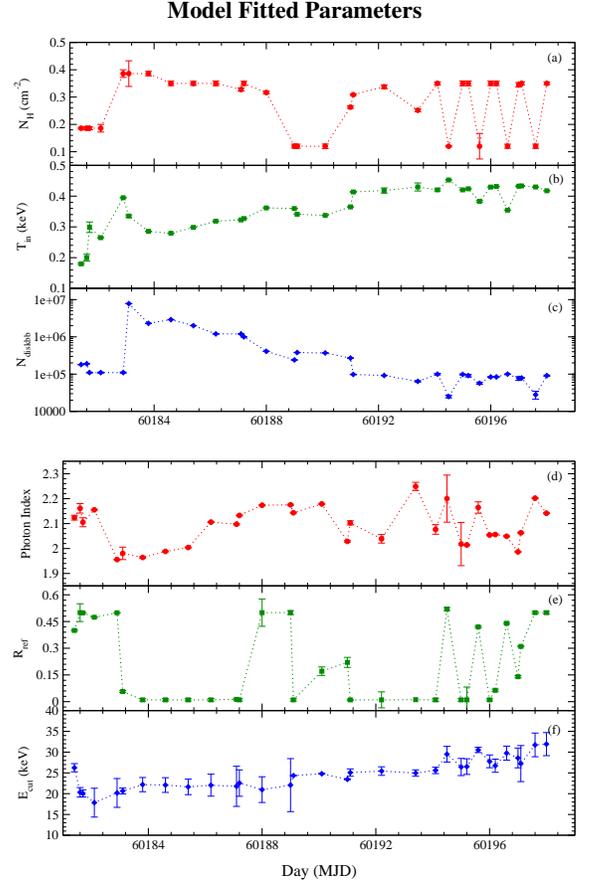

\vskip 0.2cm
\centering
\vbox{
\includegraphics[width=7.0truecm,angle=0]{mjd-nH-Tin-Norm.eps}\hskip 0.5cm
\includegraphics[width=7.0truecm,angle=0]{mjd-PhInd-Refl-Ecut.eps}}
    \caption{Variation of spectrally fitted properties (a) hydrogen column density ($n_H$), (b) inner-disk temperature ($T_{in}$ in keV), (c) diskbb normalization, (d) Photon Index ($\Gamma$), 
	     (e) reflection fraction (R$_{\rm ref}$), and (f) cutoff energy (E$_{\rm cut}$) with time.}
\end{figure}

For our overall spectral analysis with the two sets of models, we have assumed $abund = 1.0$ and Fe$_{\rm abund} = 1.0$ in the {\fontfamily{pcr}\selectfont pexrav} model with redshift ($z$) = 0. From 
the spectral fitting, we extracted various properties of the source during the first few days of the outburst. We estimated the inner-disk temperature ($T_{in}$ in keV), photon index of power-law 
($\Gamma$) which is the slope of the spectrum. We found that from MJD 60181.4 to 60190.0, there was an absorption feature at $\sim 1.65-1.95 $ keV. After this day, the {\fontfamily{pcr}\selectfont gabs} 
model was no longer needed to achieve the best fit. The $T_{in}$ was $\sim 0.19 \pm 0.02$ keV at the start of the outburst and it gradually increased to $\sim 0.43 \pm 0.03 $ keV on the last day of 
our analysis. The $\Gamma$ of the {\fontfamily{pcr}\selectfont power-law} model was $2.12 \pm 0.15$ on the first day. From the light curve in Figure 1, we notice that the flux was already very high 
at the starting date of our analysis (HXMT data was available from this date too). The value of $T_{in}$ and $\Gamma$ indicate that the source has already transitioned past its HS and could be in an 
intermediate state. As the outburst progressed,  $T_{in}$ became higher with the norm of the {\fontfamily{pcr}\selectfont diskbb} model decreased as the outburst progressed. $\Gamma$ of the 
{\fontfamily{pcr}\selectfont power-law} model was in the narrow range $\sim 2.10-2.16$ for the first few days then decreased a little and then again slowly increased to $\sim 2.20$. We found the presence of 
a broad Gaussian line, with the line energy varying between $5.2-5.8$ keV. We did not link the $\Gamma$ of the {\fontfamily{pcr}\selectfont pexrav} model to the $\Gamma$ of the {\fontfamily{pcr}\selectfont 
power-law} model, as we noticed that the spectral slope was different between low and high energies. The $\Gamma$ of the {\fontfamily{pcr}\selectfont pexrav} model varied in the range of $\sim 0.9-1.9$. The 
$E_{cut}$ varied in the range of $\sim 20-32$ keV. The $\cos(i)$ parameter varied in the range $0.05-0.1$ for most of the exposures except for the first two, where it was $\sim$ 0.5, and 0.2, respectively. 
This could be due to the degeneracy of the parameter space. The column density of H ($N_H$) varied in the range $(0.12 \pm 0.02 - 0.39 \pm 0.08)\times10^{22}$ cm$^{-2}$ during the entire duration of this 
analysis. In Figure 7(a-f), we show the variations of the spectral model fitted parameters with time.

\subsection{Correlation between spectral and timing properties}

\begin{figure}[!h]
\vskip 0.2cm
\centering
\vbox{
\includegraphics[width=7.0truecm,angle=0]{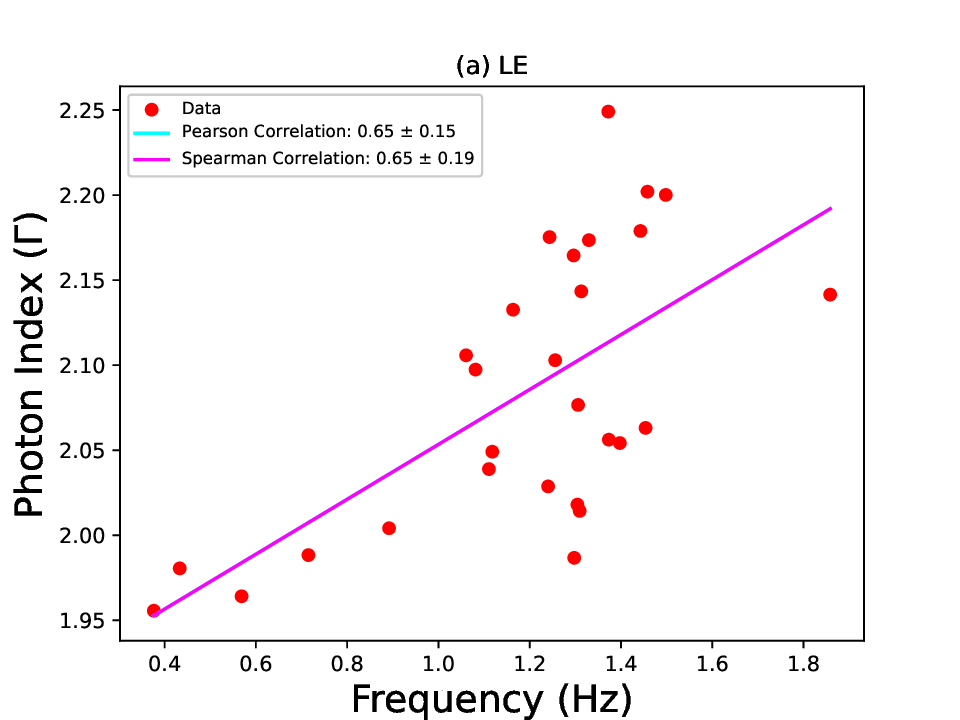}
\hskip 0.2cm
\includegraphics[width=7.0truecm,angle=0]{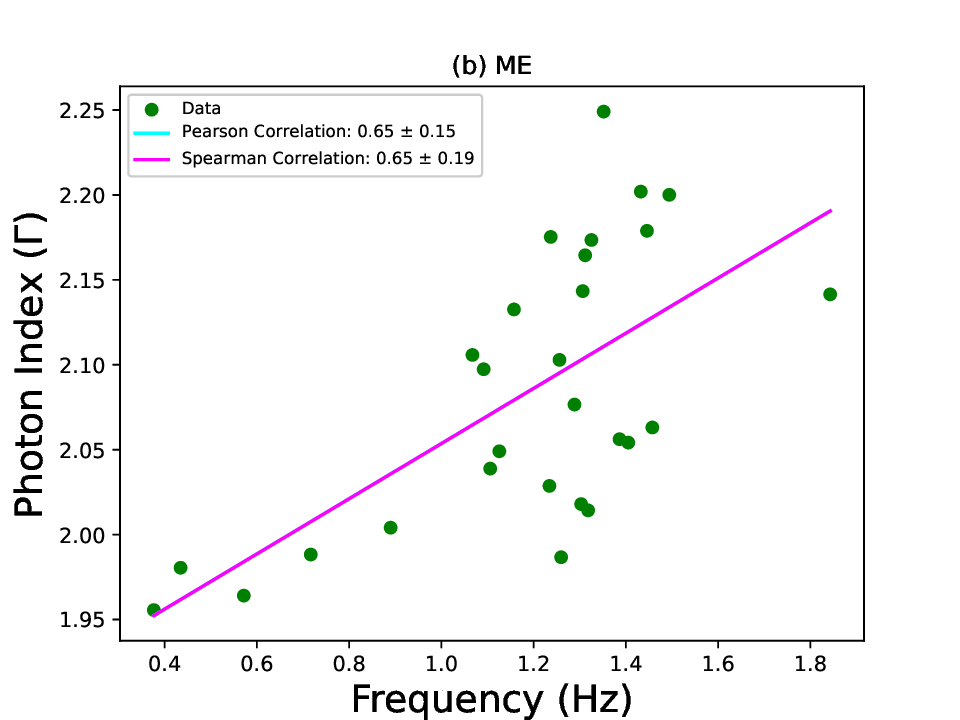}\hskip 0.5cm
\hskip 0.1cm
\includegraphics[width=7.0truecm,angle=0]{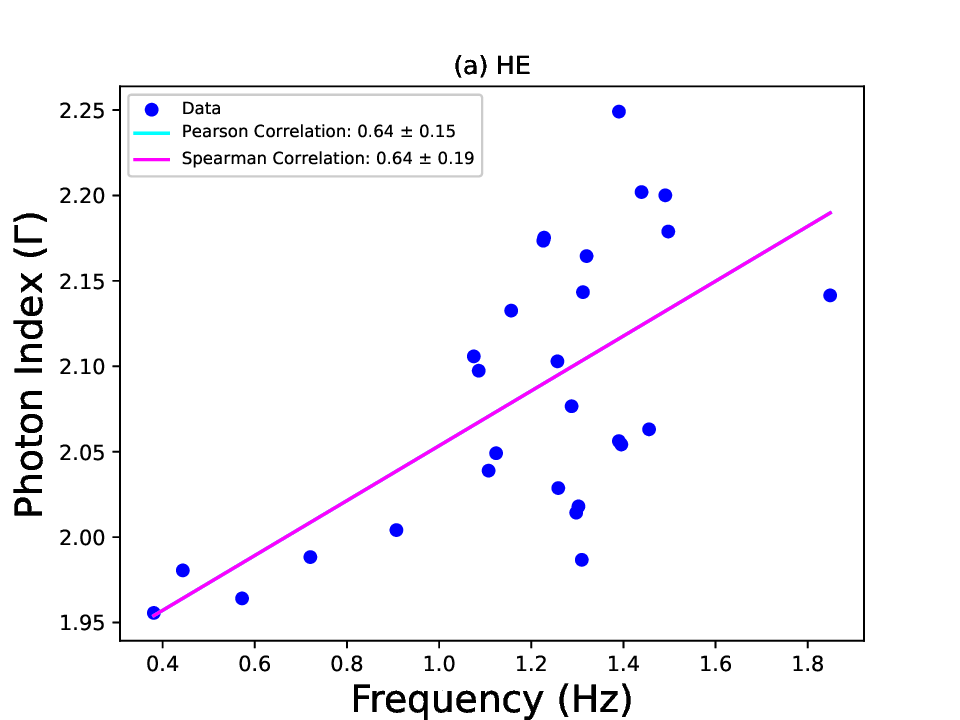}}
\caption{Variation of photon index of power-law ($\Gamma$) with QPO frequency ($\nu_{qpo}$) for (a) LE, (b) ME, and (c) HE instruments. Here, $\Gamma$ is the spectral fitted index using combined 
         LE+ME+HE data.}
\end{figure}

It has been previously seen that the QPO frequency ($\nu_{qpo}$) and $\Gamma$ show a positive correlation for some of the sources during their active outbursting phase (Vignarca et al. 2003; Shaposhnikov 
\& Titarchuk 2009; Stiele et al. 2013). We also tried to find out if there are any correlations between these two properties. In Figure 8, we show the variations of the photon index with the QPO frequency 
for the three energy bands. Considering these points, we have found a positive correlation between $\Gamma$ and $\nu_{qpo}$. Using both the Pearson-Linear and Spearman-Rank correlation methods, we found 
correlation coefficients for the three energy bands as shown in Table 2. In all of the three energy bands, we see moderately strong correlation when it comes to $\Gamma$ and $\nu_{qpo}$ 
as correlation coefficient is $\sim 0.5-0.8$. This shows the timing and spectral properties, the origin of which are the same geometric variation of the Compton corona, are moderately correlated in this 
source.

\begin{table}
\scriptsize
\caption{Correlation coefficients between the QPO frequency ($\nu_{qpo}$) and photon index of power-law ($\Gamma$).}
 \centering
 \label{tab:table2}
 \begin{tabular}{ccc}
 \hline
     &  Pearson-Linear   &   Spearman-Rank   \\
\hline
LE   &  $0.65 \pm 0.15$  &  $0.65 \pm 0.19$  \\
\hline
ME   &  $0.65 \pm 0.15$  &  $0.65 \pm 0.19$  \\
\hline
HE   &  $0.64 \pm 0.15$  &  $0.64 \pm 0.19$  \\
\hline
 \end{tabular}
\end{table}

\vskip 0.35cm
\subsection{Propagation of Shock}

\begin{figure}[!h]
  \centering
    \includegraphics[width=8.5cm]{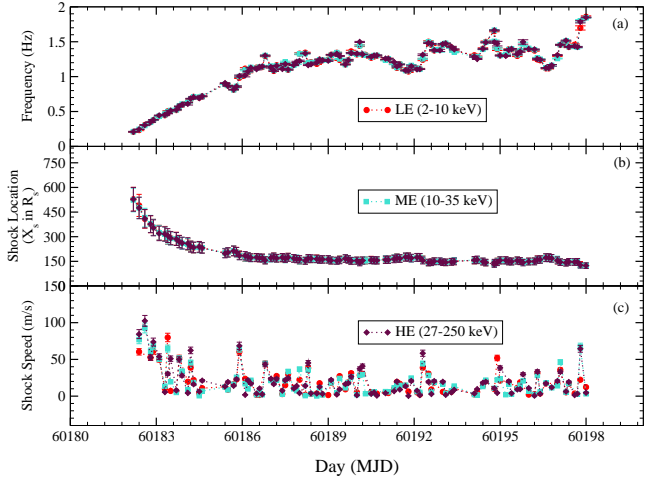}
    \caption{Variation of shock properties with time in accordance with $\nu_{qpo}$. In (a) the variation of $\nu_{qpo}$ with time and in (b) and (c) we show the variations of the shock location
            ($X_s$), and shock speed ($v$) with time.}
\end{figure}

In the TCAF paradigm, the oscillation of the shock is responsible for the origin of QPOs. When this shock front oscillates, it produces QPOs. The shock oscillation model has two types of timescales: 
the infall timescale ($t_i$) and the cooling timescale ($t_c$) and matching of these timescales gives the resonance condition responsible for the QPOs. The relation between the QPO frequency and shock location 
is given as (Molteni et al. 1996; Chakrabarti \& Manickam 2000; Chakrabarti et al. 2004),

\begin{eqnarray}
\nu_{qpo} = \frac{c^3}{2GM_{BH}} \frac{1}{RX_s (X_s -1)^{1/2}} ~\text{Hz}.
\end{eqnarray}
 
where, $c$, $G$, $M_{BH}$, $X_s$, and $R$ are the speed of light, the gravitational constant, the mass of the BH, location of the shock (in units of the Schwarzschild radius $r_s$), and the ratio of matter 
densities in post-shock to pre-shock regions ($\rho_+/\rho_-$) respectively. From our timing analysis, we extracted the information about the QPO frequency ($\nu_{qpo}$). Using the above relation, we estimated 
the shock location during the outburst. Using Eq. 1, we found that the shock was far away ($> 500~r_s, r_s = 2GM_{BH}/c^2$) from the BH at the start of the outburst (see Figure 9b). It then shifted towards the black 
hole sharply as the QPO frequency increased rapidly during the initial few days ($\sim 250~r_s$). Close to MJD 60185, the shock location started to decrease very slowly and it reached to $\sim 100~r_s$ at the end of 
our analysis period. The values of the shock location are given in Table 5.

If the time of the first exposure ID was taken to be 0, and the shock location as $X_{s0}$ and  after time $t$, if the shock is at $X_s$,  then the velocity of the movement of the shock would be given 
by the equation (Chakrabarti et al. 2005)

\begin{eqnarray}
\varv = \frac{X_{s0} - X_s}{t} ~{\text m s}^{-1}.
\end{eqnarray}

Using this relation, we have estimated the velocity of the movement of the shock. The estimated shock velocities are given in Figure 9c, which is a few tens of m/s in agreement with the previous 
estimates (Chakrabarti et al. 2005; Mondal et al. 2015). We find that the shock velocity varies within this period. In the beginning, the shock moved with a speed of $\sim 50-100$ m s$^{-1}$, when the $X_s$ was 
also larger. Then, after MJD $\sim$ 60183.7, it decreased a little bit with values $< 50$ m s$^{-1}$. Then it slowed down and was moving with a speed of $\sim 10-30$ m s$^{-1}$ with some fluctuations and continued 
till the end of the observation period.

\section{Discussion} 

The Galactic black hole Swift J1727.8-1613 went through an outburst recently in 2023. We have used {\it Insight}-HXMT data for our both spectral and timing studies from MJD 60181.4 to 60198.0. The evolution of 
the light curve makes the outburst nature of the source a fast-rise slow decay type. After the onset of the outburst, the flux reached its peak value very fast within 5 days. Using the $0.01$ sec time-binned light 
curves from the three instruments of HXMT (LE, ME, and HE), we studied the source's timing properties. Using the spectra files from these three instruments, we also studied the spectral properties of this source 
by analyzing combined LE+ME+HE spectra in the $2-150 $ keV broad energy range.

Quasi-periodic oscillation is one of the most important and common phenomena for stellar mass black holes. For this newly discovered source, we analyzed a total of 342 light curves (114 for LE, ME, and HE each).
However, in some of the exposures, light curves were not produced properly. There were a total of 92 exposures which commonly produced LE, ME, and HE light curves. We list those data in Table 3. Finally, we analyzed 
276 lightcurves. We found the presence of QPO in all of these light curves, and they showed rapid evolution in frequency in this short analysis period.  The QPO frequency ($\nu_{qpo}$) showed an increase in frequency 
within even a single day. The evolution of the QPO frequency has been discussed broadly in the result section. It showed an increasing trend from the start of our analysis. 
We found that the RMS (\%) was very high ($>13 \%$) for all the exposures. The $Q$-value was also high, though for some observations it was close to those QPOs that are types-A/B. However, the 
frequency and rms allow us to classify the QPOs as type-C, which are in agreement with Casella et al. (2005). Considering the values of RMS, $Q$-value and the type of QPO, we suggest that the source 
was in the HIMS. Recently, Yu et al. (2024) also reported the timing properties of this source for a longer period. The 
QPO frequency has varied in a much broader range in their report from 0.1 to 8 Hz. The authors also concluded the observed QPOs in the full period as type-C. During the 2009-10 outburst of the BHC XTE J1752-223 
(Chatterjee et al. 2020), we found the existence of type-C QPOs that had $Q$-values comparable to this source. During the 2016-17 outburst of the BHC GRS 1716-249 (Chatterjee et al. 2021), type-C QPOs were found 
to be present with similar RMS (\%) values. Other than these sources, Belloni et al. (2002), Casella et al. (2005), and the references therein can be seen for an analogy. Such similarities in observed QPO 
properties in different sources require physical explanations of common origin.  

Here, we would like to focus on the physical scenario that explains the origin of QPOs due to the shock oscillations in advective flows around BHs. Accretion onto BH can be explained as transonic flow with
possibility of multiple sonic points (Chakrabarti 1989). The matter from the companion need not necessarily be only Keplerian. There could be a supply of matter which has angular momentum distribution deviated from the
Keplerian one. This is the sub-Keplerian component. Such a lower angular momentum component of infalling matter gets accreted in a free-fall timescale. Due to the tug of war between the gravitational force and centrifugal
force, this matter could virtually stop at some distance from the BH and undergo shock transition, creating a post-shock region. Depending on the flow properties, the standing shocks forms (Chakrabarti 1989; Singh et al.
2022 and references therein). This approach can well describe the observed spectral and temporal properties of BHs (Debnath et al. 2014; Mondal et al. 2014, Chatterjee et al. 2020, 2021, 2023). According to Molteni 
et al. (1996), the shock can be oscillatory due to the presence of cooling. This was later shown in simulation by Garain et al. (2014). During the oscillation when the cooling timescale due to the Comptonization process 
and the heating timescale matches, QPOs originate (see Chakrabarti et al. 2015). Additionally, it has also been observed by the authors that once the QPO sets in, it gets locked for a week or more depending on the 
above condition.  

According to Chakrabarti et al. (2005), $X_s$ can be located anywhere above 10 $r_s$ depending on the flow parameters producing shocks. The shock forms far away $\sim 1000 ~r_s$ when the spectral nature of the 
ongoing outburst is hard and decreases gradually in the progressive days as cooling increases (Mondal et al. 2015). When the cooling timescale ($t_c$) and infall timescale ($t_i$) becomes comparable to each other 
(Molteni et al. 1996), an oscillation could set in the shock boundary due to resonance. This could produce variations as we observe in the light curves. This might be the case for this source regarding the 
origin of the occurrence of the QPOs. Here, from this analysis, we have found that the shock was far away ($\sim 500~r_s$) when QPO frequency ($\nu_{qpo}$) was low ($\sim 0.2$~Hz). 
The disk was far away. As the disk moves inward, cooling increases with progressive days and QPO frequency increases, as shown in Eq. 1; following inverse scaling with $X_s$. This is the case that 
we observed here. After the start, as the outburst progressed, the cooling began to take place and $X_s$ got smaller in size. As a result, $\nu_{qpo}$ increased. After a few days, the shock location decreased very 
slowly, without changing its value much, the $\nu_{qpo}$ was also increasing very slowly. We also tried to observe how the shock moved, considering the constant velocity approach, as given in Eq. 2. We have found 
that the shock speed varied in a broad range. At the start, the shock velocity was relatively higher compared to later days, while it became nearly constant with some fluctuations in later time. Such fluctuations 
can be due to the limitation of assuming constant velocity of the shock propagation. For a comparison, we refer to Chakrabarti et al. (2005). The authors showed that considering the constant velocity approximation, 
the average speed of the shock is close to 20 m s$^{-1}$, which is nearly consistent with our result. This also depends on the non-constant time difference between the exposures. However, the overall velocity profile 
agrees that the shock moved faster when $X_s$ was bigger, decreased with $X_s$, and afterward almost with constant velocity.

From the spectral properties, we have seen that the $T_{in}$ was low initially and then gradually increased. The exact opposite trend was observed for the Normalization of the {\fontfamily{pcr}\selectfont diskbb} model. 
The norm of {\fontfamily{pcr}\selectfont diskbb} model varies as 

\begin{eqnarray}
\href{https://heasarc.gsfc.nasa.gov/docs/xanadu/xspec/manual/node166.html}{N_{diskbb}}\footnote{https://heasarc.gsfc.nasa.gov/docs/xanadu/xspec/manual/node166.html} = \left(\frac{R_{in}}{D_{10}}\right)^2 \cos{i}
\end{eqnarray}

where, $R_{in}$, $D_{10}$ are the inner-disk radius of the accretion disk and the distance of the source in $10$ kpc unit. $i$ is the inclination angle of the accretion disk to the observer. Assuming that $D_{10}$ and 
$\theta$ do not change during an outburst, the relation could be shortened like $R_{in} \sim N_{diskbb}^{1/2}$. For our case, except for one exposure, the inclination varied in a narrow range of 78-86$^\circ$. Thus, as 
the $N_{diskbb}$ decreased, it means the inner edge of the disk moved closer to the BH, which agrees with the fact that $T_{in}$ increased as the disk moved closer. Furthermore, this also supports the properties we showed 
from the timing analysis that the shrinking of post-shock region with time.

The photon index of power-law ($\Gamma$) is an important property to decide the spectral nature of an ongoing outburst. This decides the slope in the spectrum, which tells whether the spectrum extends up to high 
energy or not. When an outburst stays in the hard state, the photon index of power-law ($\Gamma$) generally stays low $\sim 1.7$ (Remillard \& McClintock 2006). From the expression of power-law component ($A(E) 
\propto E^{-\Gamma}$), this suggests that the spectrum extends up to high energy. As the source gets towards intermediate states, $\Gamma$ becomes $\sim 2$ (e.g., Remillard \& McClintock 2006), even up to 3, 
sometimes. In the high soft state, $\Gamma$ reaches a value of $>3$ (Remillard \& McClintock (2006), for a review), although sources have been discovered to reach soft state with $\Gamma < 3$. According to the 
TCAF configuration (Chakrabarti \& Titarchuk 1995), as discussed earlier in the Introduction, the shock location, that produces both the change in the spectral and timing properties, 
changes in size in different spectral states. In the HS it stays very far from the source, at $\sim 1000 ~r_s$. The disk is even further away. As a result, only a few photons come in the process of radiation. As the disk 
approaches closer, the $X_s$ becomes smaller and the component from the disk starts to gradually dominate over the hard power-law component. In the intermediate states, these two components stay comparable. This time the 
shock stays at a few $100 ~r_s$. In the SS, the disk moves even closer and $X_S$ becomes very small (a few tens of $r_s$). Thus, the disk component now fully dominates, and the spectrum becomes very soft. Thereby, the location 
of the shock determines the spectral state of the outburst. Looking at the spectral parameters, and also the location of the shock, we propose that the source was already in an intermediate state at the start of our 
analysis period. As the outburst progressed, it was transitioning towards softer states. From the values of the parameter $\cos(i)$  of the {\fontfamily{pcr}\selectfont pexrav} model, we find that the inclination of the 
disk was $>60^\circ$, which makes this source a highly inclined system. This is in agreement with the recent reports of Bouchet et al. (2024), Yu et al. (2024). The hydrogen column density ($N_H$) showed a variation, 
which is common for Galactic black hole binaries (Eckersall et al. 2017; Chatterjee et al. 2023) unless the system is heavily obscured, which has recently been found for the newly discovered BH Swift J151857.0-572147 
(Mondal et al. 2024), Swift J1658.2-4242 (Mondal \& Jithesh 2023 and references therein). The $N_H$ has a variation in the range of $(0.12 \pm 0.02 - 0.39 \pm 0.08)\times10^{22}$ cm$^{-2}$.

The QPO frequency can be produced from the oscillation of the Compton corona, which can also be the origin of the hard power-law photons that extend the spectrum to higher energies (e.g., Lee \& Miller 1998; 
Kumar \& Mishra 2014; Karpouzas et al. 2020; Liu et al. 2022; Bellavita et al. 2022; etc.). Therefore, these two features might be interlinked or correlated. As described in the introduction, the oscillation of the CENBOL can be the 
mechanism producing the QPOs. This CENBOL is also responsible for producing the power-law part of the spectrum. The change in the size of the shock location gives the change of spectral states as well as the frequency 
of the QPOs. We found such correlations for the source Swift J1727.8-1613 during its 2023 outburst.

\section{Summary and Conclusions}

We have studied the timing and spectral properties of the BHC Swift J1727.8-1613 during its recent outburst in 2023. We show the evolution of the light curve along with its hardness ratio for the entire duration of the 
outburst using archival MAXI/GSC data. Studying the entire evolution of the outburst was not the scope of this paper. Using the {\it Insight}-HXMT data, we chose the first ten observation IDs for our analysis. We use all the 
exposures from those observation IDs and selective exposures for timing and spectral analysis, respectively. Using $0.01 ~s$ time-binned light curves from all three instruments of HXMT, i.e., LE, ME, and HE, we studied
the QPO properties by producing a power density spectrum. The QPO properties were extracted with the use of {\fontfamily{pcr}\selectfont Lorentzian} model. For spectral analysis, we use {\fontfamily{pcr}\selectfont LE 
+ ME + HE} spectrum files in the broad $2-150 $ keV energy band. We found that the models i) {\fontfamily{pcr}\selectfont constant*tbabs*gabs*(diskbb + power-law + gaussian + pexrav)} (for those for which 
{\fontfamily{pcr}\selectfont gabs} was required), ii) {\fontfamily{pcr}\selectfont constant*tbabs*(diskbb + power-law + gaussian + pexrav)} fits the spectra best. From our analysis, we conclude that:

(i) Quasi-periodic oscillations were present during the entire period of our analysis, which showed evolution in QPO frequency from $0.21 \pm 0.01$ to $1.86 \pm 0.01$ Hz within this analysis period. 

(ii) The values of the $Q$-factor and $RMS$ best represent the QPOs as type-C.

(iii) Our study infers that the QPOs might have originated due to the resonance oscillations of the shocks. As the shock moved inwards (outwards), the QPO frequency increased (decreased).

(iv) With the shock moving inwards, the spectral nature of the source was transitioning towards the softer states, after it started in the intermediate state at the onset of the analysis period.

(v) The spectral and temporal properties correlate to each other which makes our claim stronger about the origin of QPOs.

(vi) This source has a high inclination.

(vii) The hydrogen column density varied in the range of $(0.12 \pm 0.02 - 0.39 \pm 0.08)\times10^{22}$ cm$^{-2}$., which is quite common for interstellar absorption of Galactic black holes.

\section{Data Availability}

This work has made use of public data from several satellite/instrument archives and has made use of software from the HEASARC, which is developed and monitored by the Astrophysics Science Division at NASA/GSFC and the High
Energy Astrophysics Division of the Smithsonian Astrophysical Observatory. This work made use of the data from the {\it Insight}-HXMT mission, a project funded by the China National Space Administration (CNSA) and the Chinese Academy 
of Sciences (CAS).

\section{Acknowledgements}
We thank the scientific editor and the anonymous referee(s) for their detailed comments and insightful suggestions that improved the quality of the paper.
KC acknowledges support from the SWIFAR postdoctoral fund of Yunnan University. SM acknowledges the Ramanujan Fellowship (\# RJF/2020/000113) by SERB/ANRF-DST, Govt. of India for this research. CBS is supported by the National 
Natural Science Foundation of China under grant no. 12073021.


\addtolength{\tabcolsep}{-1.5pt}
\hspace{-50.0cm}
\begin{longtable}{|c|c|c|c|c|c|}
  \caption{Start and Stop Time of all the HXMT Exposures}\label{tab:long} \\
 \hline
     Exposure ID              &    UT Start     &      MJD Start    &     UT Stop     &     MJD Stop   &     Average MJD     \\
          (1)                 &       (2)       &         (3)       &       (4)       &        (5)     &        (6)          \\
\hline
\endfirsthead
\multicolumn{6}{c}
{{\bfseries \tablename\ \thetable{} -- continued from previous page}} \\
\hline
     Exposure ID              &    UT Start     &      MJD Start    &     UT Stop     &     MJD Stop   &     Average MJD     \\
          (1)                 &       (2)       &         (3)       &       (4)       &        (5)     &        (6)          \\
\hline
\endhead

\hline \multicolumn{6}{|r|}{Continued on next page} \\ \hline
\endfoot

\endlastfoot
P061433800101-20230825-01-01* &   2023-08-25    &      60181.34     &    2023-08-25   &     60181.49   &       60181.42       \\  
P061433800102-20230825-01-01* &   2023-08-25    &      60181.49     &    2023-08-25   &     60181.63   &       60181.66       \\  
P061433800103-20230825-01-01* &   2023-08-25    &      60181.63     &    2023-08-25   &     60181.76   &       60181.70       \\  
P061433800104-20230825-01-01  &   2023-08-25    &      60181.76     &    2023-08-25   &     60181.89   &       60181.83       \\  
P061433800105-20230825-01-01  &   2023-08-25    &      60181.89     &    2023-08-26   &     60182.02   &       60182.06       \\  
P061433800106-20230826-02-01* &   2023-08-26    &      60182.02     &    2023-08-26   &     60182.15   &       60182.19       \\  
P061433800107-20230826-02-01  &   2023-08-26    &      60182.15     &    2023-08-26   &     60182.28   &       60182.22       \\  
P061433800108-20230826-02-01  &   2023-08-26    &      60182.28     &    2023-08-26   &     60182.42   &       60182.45       \\  
P061433800110-20230826-02-01  &   2023-08-26    &      60182.55     &    2023-08-26   &     60182.68   &       60182.62       \\  
P061433800111-20230826-02-01  &   2023-08-26    &      60182.68     &    2023-08-26   &     60182.81   &       60182.85       \\  
P061433800112-20230826-02-01* &   2023-08-26    &      60182.81     &    2023-08-26   &     60182.95   &       60182.99       \\  
P061433800113-20230826-02-01  &   2023-08-26    &      60182.95     &    2023-08-27   &     60183.05   &       60183.01       \\  
P061433800201-20230827-01-01* &   2023-08-27    &      60183.05     &    2023-08-27   &     60183.20   &       60183.13       \\  
P061433800202-20230827-01-01  &   2023-08-27    &      60183.20     &    2023-08-27   &     60183.34   &       60183.37       \\  
P061433800203-20230827-01-01  &   2023-08-27    &      60183.34     &    2023-08-27   &     60183.48   &       60183.41       \\  
P061433800204-20230827-01-01  &   2023-08-27    &      60183.48     &    2023-08-27   &     60183.61   &       60183.55       \\  
P061433800205-20230827-01-01  &   2023-08-27    &      60183.61     &    2023-08-27   &     60183.74   &       60183.78       \\  
P061433800206-20230827-01-01* &   2023-08-27    &      60183.74     &    2023-08-27   &     60183.87   &       60183.81       \\  
P061433800207-20230827-01-01  &   2023-08-27    &      60183.87     &    2023-08-28   &     60184.00   &       60183.94       \\  
P061433800208-20230828-02-01  &   2023-08-28    &      60184.00     &    2023-08-28   &     60184.14   &       60184.17       \\  
P061433800209-20230828-02-01  &   2023-08-28    &      60184.14     &    2023-08-28   &     60184.26   &       60184.20       \\  
P061433800210-20230828-02-01  &   2023-08-28    &      60184.26     &    2023-08-28   &     60184.40   &       60184.34       \\  
P061433800211-20230828-02-01  &   2023-08-28    &      60184.40     &    2023-08-28   &     60184.53   &       60184.57       \\  
P061433800212-20230828-02-01* &   2023-08-28    &      60184.53     &    2023-08-28   &     60184.64   &       60184.69       \\  
P061433800301-20230829-01-01* &   2023-08-29    &      60185.30     &    2023-08-29   &     60185.46   &       60185.48       \\  
P061433800302-20230829-01-01  &   2023-08-29    &      60185.46     &    2023-08-29   &     60185.59   &       60185.53       \\  
P061433800303-20230829-01-01  &   2023-08-29    &      60185.59     &    2023-08-29   &     60185.72   &       60185.76       \\  
P061433800304-20230829-01-01  &   2023-08-29    &      60185.72     &    2023-08-29   &     60185.85   &       60185.89       \\  
P061433800305-20230829-01-01  &   2023-08-29    &      60185.85     &    2023-08-29   &     60185.99   &       60185.92       \\  
P061433800306-20230829-01-01  &   2023-08-29    &      60185.99     &    2023-08-30   &     60186.12   &       60186.16       \\  
P061433800307-20230830-02-01* &   2023-08-30    &      60186.12     &    2023-08-30   &     60186.24   &       60186.29       \\  
P061433800308-20230830-02-01  &   2023-08-30    &      60186.24     &    2023-08-30   &     60186.38   &       60186.32       \\  
P061433800309-20230830-02-01  &   2023-08-30    &      60186.38     &    2023-08-30   &     60186.51   &       60186.55       \\  
P061433800310-20230830-02-01  &   2023-08-30    &      60186.51     &    2023-08-30   &     60186.65   &       60186.69       \\  
P061433800311-20230830-02-01  &   2023-08-30    &      60186.65     &    2023-08-30   &     60186.78   &       60186.72       \\  
P061433800312-20230830-02-01  &   2023-08-30    &      60186.78     &    2023-08-30   &     60186.91   &       60186.85       \\  
P061433800313-20230830-02-01  &   2023-08-30    &      60186.91     &    2023-08-31   &     60187.04   &       60187.08       \\  
P061433800314-20230831-03-01* &   2023-08-31    &      60187.04     &    2023-08-31   &     60187.15   &       60187.10       \\  
P061433800401-20230831-01-01* &   2023-08-31    &      60187.15     &    2023-08-31   &     60187.30   &       60187.23       \\  
P061433800402-20230831-01-01  &   2023-08-31    &      60187.30     &    2023-08-31   &     60187.44   &       60187.48       \\  
P061433800403-20230831-01-01  &   2023-08-31    &      60187.44     &    2023-08-31   &     60187.57   &       60187.51       \\  
P061433800404-20230831-01-01  &   2023-08-31    &      60187.57     &    2023-08-31   &     60187.70   &       60187.64       \\  
P061433800405-20230831-01-01  &   2023-08-31    &      60187.70     &    2023-08-31   &     60187.84   &       60187.87       \\  
P061433800406-20230831-01-01  &   2023-08-31    &      60187.84     &    2023-08-31   &     60187.97   &       60187.91       \\  
P061433800407-20230831-01-01* &   2023-08-31    &      60187.97     &    2023-09-01   &     60188.10   &       60188.04       \\  
P061433800408-20230901-02-01  &   2023-09-01    &      60188.10     &    2023-09-01   &     60188.22   &       60188.27       \\  
P061433800409-20230901-02-01  &   2023-09-01    &      60188.22     &    2023-09-01   &     60188.36   &       60188.30       \\  
P061433800410-20230901-02-01  &   2023-09-01    &      60188.36     &    2023-09-01   &     60188.50   &       60188.43       \\  
P061433800411-20230901-02-01  &   2023-09-01    &      60188.50     &    2023-09-01   &     60188.63   &       60188.67       \\  
P061433800412-20230901-02-01  &   2023-09-01    &      60188.63     &    2023-09-01   &     60188.76   &       60188.70       \\  
P061433800413-20230901-02-01  &   2023-09-01    &      60188.76     &    2023-09-01   &     60188.89   &       60188.83       \\  
P061433800414-20230901-02-01* &   2023-09-01    &      60188.89     &    2023-09-02   &     60189.07   &       60189.09       \\  
P061433800501-20230902-01-01* &   2023-09-02    &      60189.07     &    2023-09-02   &     60189.22   &       60189.15       \\  
P061433800502-20230902-01-01  &   2023-09-02    &      60189.22     &    2023-09-02   &     60189.35   &       60189.39       \\  
P061433800503-20230902-01-01  &   2023-09-02    &      60189.35     &    2023-09-02   &     60189.49   &       60189.43       \\  
P061433800504-20230902-01-01  &   2023-09-02    &      60189.49     &    2023-09-02   &     60189.62   &       60189.66       \\  
P061433800505-20230902-01-01  &   2023-09-02    &      60189.62     &    2023-09-02   &     60189.75   &       60189.79       \\  
P061433800506-20230902-01-01  &   2023-09-02    &      60189.75     &    2023-09-02   &     60189.88   &       60189.82       \\  
P061433800507-20230902-01-01  &   2023-09-02    &      60189.88     &    2023-09-03   &     60190.02   &       60190.05       \\  
P061433800508-20230903-02-01* &   2023-09-03    &      60190.02     &    2023-09-03   &     60190.15   &       60190.19       \\  
P061433800509-20230903-02-01  &   2023-09-03    &      60190.15     &    2023-09-03   &     60190.28   &       60190.22       \\  
P061433800510-20230903-02-01  &   2023-09-03    &      60190.28     &    2023-09-03   &     60190.41   &       60190.35       \\  
P061433800511-20230903-02-01  &   2023-09-03    &      60190.41     &    2023-09-03   &     60190.54   &       60190.58       \\  
P061433800512-20230903-02-01  &   2023-09-03    &      60190.54     &    2023-09-03   &     60190.68   &       60190.62       \\  
P061433800513-20230903-02-01  &   2023-09-03    &      60190.68     &    2023-09-03   &     60190.81   &       60190.75       \\  
P061433800514-20230903-02-01  &   2023-09-03    &      60190.81     &    2023-09-03   &     60190.94   &       60190.98       \\  
P061433800515-20230903-02-01* &   2023-09-03    &      60190.94     &    2023-09-04   &     60191.06   &       60191.00       \\  
P061433800601-20230904-01-01* &   2023-09-04    &      60191.06     &    2023-09-04   &     60191.20   &       60191.13       \\  
P061433800602-20230904-01-01  &   2023-09-04    &      60191.20     &    2023-09-04   &     60191.34   &       60191.37       \\  
P061433800603-20230904-01-01  &   2023-09-04    &      60191.34     &    2023-09-04   &     60191.47   &       60191.41       \\  
P061433800604-20230904-01-01  &   2023-09-04    &      60191.47     &    2023-09-04   &     60191.60   &       60191.54       \\  
P061433800605-20230904-01-01  &   2023-09-04    &      60191.60     &    2023-09-04   &     60191.73   &       60191.77       \\  
P061433800606-20230904-01-01  &   2023-09-04    &      60191.73     &    2023-09-04   &     60191.87   &       60191.80       \\  
P061433800607-20230904-01-01  &   2023-09-04    &      60191.87     &    2023-09-05   &     60192.00   &       60191.94       \\  
P061433800608-20230905-02-01  &   2023-09-05    &      60192.00     &    2023-09-05   &     60192.13   &       60192.17       \\  
P061433800609-20230905-02-01* &   2023-09-05    &      60192.13     &    2023-09-05   &     60192.26   &       60192.20       \\  
P061433800610-20230905-02-01  &   2023-09-05    &      60192.26     &    2023-09-05   &     60192.39   &       60192.33       \\  
P061433800611-20230905-02-01  &   2023-09-05    &      60192.39     &    2023-09-05   &     60192.53   &       60192.57       \\  
P061433800612-20230905-02-01  &   2023-09-05    &      60192.53     &    2023-09-05   &     60192.66   &       60192.60       \\  
P061433800613-20230905-02-01  &   2023-09-05    &      60192.66     &    2023-09-05   &     60192.79   &       60192.73       \\  
P061433800614-20230905-02-01  &   2023-09-05    &      60192.79     &    2023-09-05   &     60192.92   &       60192.96       \\  
P061433800615-20230905-02-01  &   2023-09-05    &      60192.92     &    2023-09-06   &     60193.06   &       60193.09       \\  
P061433800616-20230906-03-01  &   2023-09-06    &      60193.06     &    2023-09-06   &     60193.19   &       60193.13       \\  
P061433800617-20230906-03-01  &   2023-09-06    &      60193.19     &    2023-09-06   &     60193.32   &       60193.36       \\  
P061433800618-20230906-03-01* &   2023-09-06    &      60193.32     &    2023-09-06   &     60193.44   &       60193.48       \\  
P061433800801-20230907-01-01* &   2023-09-07    &      60194.03     &    2023-09-07   &     60194.17   &       60194.10       \\  
P061433800802-20230907-01-01  &   2023-09-07    &      60194.17     &    2023-09-07   &     60194.31   &       60194.24       \\  
P061433800803-20230907-01-01  &   2023-09-07    &      60194.31     &    2023-09-07   &     60194.44   &       60194.48       \\  
P061433800804-20230907-01-01* &   2023-09-07    &      60194.44     &    2023-09-07   &     60194.58   &       60194.51       \\  
P061433800805-20230907-01-01  &   2023-09-07    &      60194.71     &    2023-09-07   &     60194.84   &       60194.88       \\  
P061433800806-20230907-01-01  &   2023-09-07    &      60194.71     &    2023-09-07   &     60194.84   &       60194.88       \\  
P061433800807-20230907-01-01  &   2023-09-07    &      60194.84     &    2023-09-07   &     60194.97   &       60194.91       \\  
P061433800808-20230907-01-01* &   2023-09-07    &      60194.97     &    2023-09-08   &     60195.09   &       60195.03       \\  
P061433800901-20230908-01-01* &   2023-09-08    &      60195.09     &    2023-09-08   &     60195.23   &       60195.26       \\  
P061433800902-20230908-01-01  &   2023-09-08    &      60195.23     &    2023-09-08   &     60195.37   &       60195.30       \\  
P061433800903-20230908-01-01  &   2023-09-08    &      60195.37     &    2023-09-08   &     60195.50   &       60195.44       \\  
P061433800904-20230908-01-01* &   2023-09-08    &      60195.50     &    2023-09-08   &     60195.63   &       60195.67       \\  
P061433800905-20230908-01-01  &   2023-09-08    &      60195.63     &    2023-09-08   &     60195.76   &       60195.70       \\  
P061433800906-20230908-01-01  &   2023-09-08    &      60195.76     &    2023-09-08   &     60195.90   &       60195.84       \\  
P061433800907-20230908-01-01* &   2023-09-08    &      60195.90     &    2023-09-09   &     60196.08   &       60196.09       \\  
P061433801001-20230909-01-01* &   2023-09-09    &      60196.08     &    2023-09-09   &     60196.22   &       60196.25       \\  
P061433801002-20230909-01-01  &   2023-09-09    &      60196.22     &    2023-09-09   &     60196.36   &       60196.39       \\  
P061433801003-20230909-01-01  &   2023-09-09    &      60196.36     &    2023-09-09   &     60196.49   &       60196.43       \\  
P061433801004-20230909-01-01* &   2023-09-09    &      60196.49     &    2023-09-09   &     60196.62   &       60196.66       \\  
P061433801005-20230909-01-01  &   2023-09-09    &      60196.62     &    2023-09-09   &     60196.76   &       60196.79       \\  
P061433801006-20230909-01-01  &   2023-09-09    &      60196.76     &    2023-09-09   &     60196.89   &       60196.83       \\  
P061433801007-20230909-01-01* &   2023-09-09    &      60196.89     &    2023-09-10   &     60197.07   &       60197.08       \\  
P061433801101-20230910-01-01* &   2023-09-10    &      60197.07     &    2023-09-10   &     60197.21   &       60197.15       \\  
P061433801102-20230910-01-01  &   2023-09-10    &      60197.21     &    2023-09-10   &     60197.35   &       60197.39       \\  
P061433801103-20230910-01-01  &   2023-09-10    &      60197.35     &    2023-09-10   &     60197.48   &       60197.42       \\  
P061433801104-20230910-01-01* &   2023-09-10    &      60197.48     &    2023-09-10   &     60197.62   &       60197.65       \\  
P061433801105-20230910-01-01  &   2023-09-10    &      60197.62     &    2023-09-10   &     60197.75   &       60197.79       \\  
P061433801106-20230910-01-01  &   2023-09-10    &      60197.75     &    2023-09-10   &     60197.88   &       60197.82       \\  
P061433801107-20230910-01-01* &   2023-09-10    &      60197.88     &    2023-09-11   &     60198.06   &       60198.08       \\  
\hline
\end{longtable}
\leftline{{\large Column 1 represents the Exposure IDs, taken for this complete analysis.}}
\leftline{{\large Column 2 and 4 represent the time (UT) of start and end of those exposures.}}
\leftline{{\large Column 3 and 5 represent the start and end MJDs of those exposures respectively.}}
\leftline{{\large Column 6 represents the average MJD for those exposure IDs.}}

\newpage
\addtolength{\tabcolsep}{-1.16pt}
\begin{longtable}{|c|c|c|c|c|c|c|c|c|c|c|c|c|}
\caption{Properties estimated using timing analysis} \label{tab:long} \\
\hline
Time      &              \multicolumn{3}{|c|}{QPO Frequency (Hz)}               &                      \multicolumn{3}{|c|}{Q-Value}                     &                      \multicolumn{3}{|c|}{RMS}                        &        \multicolumn{3}{|c|}{Shock Location ($X_s$)}            \\
\hline  
 (MJD)     &         LE           &            ME        &            HE         &          LE          &            ME          &           HE           &           LE          &            ME         &          HE           &      LE          &         ME           &          HE         \\   
\hline                                                                                                                                          
  (1)      &        (2)           &           (3)        &           (4)         &          (5)         &            (6)         &           (7)          &           (8)         &           (9)         &         (10)          &           (11)         &        (12)          &         (13)  \\ 
  \hline
\endfirsthead

\multicolumn{13}{c}
{{\bfseries \tablename\ \thetable{} -- continued from previous page}} \\
\hline 
Time      &              \multicolumn{3}{|c|}{QPO Frequency (Hz)}               &                      \multicolumn{3}{|c|}{Q-Value}                     &                      \multicolumn{3}{|c|}{RMS}                        &        \multicolumn{3}{|c|}{Shock Location ($X_s$)}            \\
\hline  
 (MJD)     &         LE           &            ME        &            HE         &          LE          &            ME          &           HE           &           LE          &            ME         &          HE           &       LE          &         ME           &          HE         \\   
\hline                                                                                                                                          
  (1)      &        (2)           &           (3)        &           (4)         &          (5)         &            (6)         &           (7)          &           (8)         &           (9)         &         (10)          &      (11)         &        (12)          &         (13)        \\  
  \hline

\endhead

\hline \multicolumn{13}{|r|}{{Continued on next page}} \\ \hline
\endfoot

\hline \hline
\endlastfoot
 2.19   &   $ 0.21 \pm{0.01} $ &  $ 0.21 \pm{0.01} $  &  $ 0.21 \pm{0.01} $   &   $ 2.2 \pm{0.5} $ &   $ 3.6 \pm{0.6} $   &   $ 2.5 \pm{0.6} $   &   $ 18.0 \pm{2.8} $ &  $ 17.8 \pm{2.2} $  &  $ 23.3 \pm{3.3} $  &   $  528 \pm{72} $ & $ 523 \pm{72} $  & $ 528 \pm{72} $ \\
 2.39   &   $ 0.23 \pm{0.01} $ &  $ 0.24 \pm{0.01} $  &  $ 0.24 \pm{0.01} $   &   $ 2.6 \pm{0.7} $ &   $ 4.4 \pm{0.7} $   &   $ 5.0 \pm{0.7} $   &   $ 17.5 \pm{3.1} $ &  $ 16.3 \pm{1.8} $  &  $ 19.7 \pm{2.0} $  &   $  490 \pm{67} $ & $ 475 \pm{65} $  & $ 475 \pm{65} $ \\ 
 2.60   &   $ 0.31 \pm{0.01} $ &  $ 0.30 \pm{0.01} $  &  $ 0.31 \pm{0.01} $   &   $ 4.2 \pm{0.7} $ &   $ 3.2 \pm{0.5} $   &   $ 4.3 \pm{0.6} $   &   $ 16.3 \pm{2.1} $ &  $ 18.2 \pm{2.0} $  &  $ 21.7 \pm{2.3} $  &   $  408 \pm{56} $ & $ 416 \pm{57} $  & $ 410 \pm{56} $ \\ 
 2.80   &   $ 0.35 \pm{0.01} $ &  $ 0.34 \pm{0.01} $  &  $ 0.35 \pm{0.01} $   &   $ 8.4 \pm{1.6} $ &   $ 5.5 \pm{0.8} $   &   $ 6.8 \pm{1.4} $   &   $ 14.2 \pm{2.0} $ &  $ 17.4 \pm{1.8} $  &  $ 20.3 \pm{2.8} $  &   $  376 \pm{52} $ & $ 378 \pm{52} $  & $ 377 \pm{52} $ \\ 
 2.89   &   $ 0.38 \pm{0.01} $ &  $ 0.38 \pm{0.01} $  &  $ 0.38 \pm{0.01} $   &   $ 4.7 \pm{1.0} $ &   $ 5.2 \pm{0.9} $   &   $ 4.8 \pm{0.7} $   &   $ 16.1 \pm{2.6} $ &  $ 17.9 \pm{2.4} $  &  $ 21.9 \pm{2.5} $  &   $  357 \pm{49} $ & $ 357 \pm{49} $  & $ 354 \pm{49} $ \\ 
 3.10   &   $ 0.43 \pm{0.01} $ &  $ 0.43 \pm{0.01} $  &  $ 0.44 \pm{0.01} $   &   $ 4.8 \pm{1.0} $ &   $ 4.6 \pm{0.7} $   &   $ 6.5 \pm{0.7} $   &   $ 16.8 \pm{2.4} $ &  $ 18.1 \pm{2.1} $  &  $ 19.7 \pm{1.4} $  &   $  325 \pm{45} $ & $ 324 \pm{44} $  & $ 320 \pm{44} $ \\ 
 3.30   &   $ 0.45 \pm{0.01} $ &  $ 0.45 \pm{0.01} $  &  $ 0.45 \pm{0.01} $   &   $ 5.1 \pm{1.0} $ &   $ 7.0 \pm{0.4} $   &   $ 6.8 \pm{0.4} $   &   $ 16.9 \pm{2.4} $ &  $ 18.7 \pm{1.2} $  &  $ 22.2 \pm{1.4} $  &   $  319 \pm{44} $ & $ 316 \pm{43} $  & $ 316 \pm{43} $ \\ 
 3.39   &   $ 0.50 \pm{0.01} $ &  $ 0.50 \pm{0.01} $  &  $ 0.47 \pm{0.01} $   &   $ 4.0 \pm{0.7} $ &   $ 4.4 \pm{0.6} $   &   $ 6.3 \pm{1.2} $   &   $ 17.8 \pm{2.3} $ &  $ 19.5 \pm{2.0} $  &  $ 21.8 \pm{3.0} $  &   $  294 \pm{40} $ & $ 296 \pm{41} $  & $ 307 \pm{42} $ \\ 
 3.50   &   $ 0.51 \pm{0.01} $ &  $ 0.51 \pm{0.01} $  &  $ 0.51 \pm{0.01} $   &   $ 5.8 \pm{1.1} $ &   $ 6.0 \pm{1.0} $   &   $ 5.3 \pm{0.9} $   &   $ 16.9 \pm{2.2} $ &  $ 18.4 \pm{2.1} $  &  $ 23.0 \pm{3.2} $  &   $  292 \pm{40} $ & $ 290 \pm{40} $  & $ 291 \pm{40} $ \\ 
 3.69   &   $ 0.52 \pm{0.01} $ &  $ 0.52 \pm{0.01} $  &  $ 0.52 \pm{0.01} $   &   $ 6.1 \pm{0.9} $ &   $ 5.6 \pm{0.7} $   &   $ 5.7 \pm{0.6} $   &   $ 16.5 \pm{1.9} $ &  $ 18.9 \pm{1.6} $  &  $ 22.7 \pm{1.8} $  &   $  288 \pm{40} $ & $ 286 \pm{39} $  & $ 286 \pm{39} $ \\ 
 3.80   &   $ 0.57 \pm{0.01} $ &  $ 0.57 \pm{0.01} $  &  $ 0.57 \pm{0.01} $   &   $ 4.7 \pm{0.7} $ &   $ 6.7 \pm{0.9} $   &   $ 6.2 \pm{0.8} $   &   $ 17.2 \pm{2.1} $ &  $ 18.7 \pm{1.8} $  &  $ 22.6 \pm{2.2} $  &   $  271 \pm{37} $ & $ 270 \pm{37} $  & $ 270 \pm{37} $ \\ 
 4.10   &   $ 0.63 \pm{0.01} $ &  $ 0.63 \pm{0.01} $  &  $ 0.61 \pm{0.02} $   &   $ 6.4 \pm{2.7} $ &   $ 8.7 \pm{2.4} $   &   $ 3.0 \pm{0.9} $   &   $ 15.5 \pm{5.0} $ &  $ 17.5 \pm{3.8} $  &  $ 19.3 \pm{4.0} $  &   $  252 \pm{35} $ & $ 254 \pm{35} $  & $ 258 \pm{36} $ \\ 
 4.19   &   $ 0.68 \pm{0.01} $ &  $ 0.68 \pm{0.01} $  &  $ 0.68 \pm{0.01} $   &   $ 7.1 \pm{1.0} $ &   $ 6.5 \pm{0.8} $   &   $ 7.4 \pm{1.1} $   &   $ 16.2 \pm{1.6} $ &  $ 19.3 \pm{1.7} $  &  $ 22.3 \pm{2.4} $  &   $  240 \pm{33} $ & $ 240 \pm{33} $  & $ 239 \pm{33} $ \\ 
 4.30   &   $ 0.71 \pm{0.01} $ &  $ 0.71 \pm{0.01} $  &  $ 0.71 \pm{0.01} $   &   $ 6.2 \pm{0.8} $ &   $ 7.2 \pm{0.7} $   &   $ 6.5 \pm{0.6} $   &   $ 16.8 \pm{1.6} $ &  $ 18.8 \pm{1.4} $  &  $ 22.7 \pm{1.6} $  &   $  234 \pm{32} $ & $ 234 \pm{32} $  & $ 234 \pm{32} $ \\ 
 4.50   &   $ 0.70 \pm{0.01} $ &  $ 0.71 \pm{0.01} $  &  $ 0.69 \pm{0.01} $   &   $ 5.9 \pm{0.2} $ &   $ 4.1 \pm{0.4} $   &   $ 5.9 \pm{0.5} $   &   $ 18.0 \pm{1.1} $ &  $ 19.8 \pm{1.5} $  &  $ 21.9 \pm{2.5} $  &   $  236 \pm{32} $ & $ 234 \pm{32} $  & $ 238 \pm{33} $ \\ 
 4.60   &   $ 0.72 \pm{0.01} $ &  $ 0.72 \pm{0.01} $  &  $ 0.72 \pm{0.01} $   &   $ 6.7 \pm{1.0} $ &   $ 7.2 \pm{0.7} $   &   $ 6.8 \pm{0.8} $   &   $ 16.6 \pm{1.9} $ &  $ 19.0 \pm{1.6} $  &  $ 22.3 \pm{2.1} $  &   $  233 \pm{32} $ & $ 232 \pm{32} $  & $ 231 \pm{32} $ \\ 
 5.39   &   $ 0.89 \pm{0.01} $ &  $ 0.89 \pm{0.01} $  &  $ 0.91 \pm{0.01} $   &   $ 7.7 \pm{0.2} $ &   $ 6.3 \pm{0.5} $   &   $ 5.3 \pm{0.5} $   &   $ 17.0 \pm{0.8} $ &  $ 19.7 \pm{1.2} $  &  $ 22.9 \pm{1.8} $  &   $  201 \pm{28} $ & $ 201 \pm{28} $  & $ 198 \pm{27} $ \\ 
 5.50   &   $ 0.87 \pm{0.01} $ &  $ 0.87 \pm{0.01} $  &  $ 0.87 \pm{0.01} $   &   $ 7.1 \pm{0.8} $ &   $ 7.2 \pm{0.7} $   &   $ 6.8 \pm{0.8} $   &   $ 16.0 \pm{1.5} $ &  $ 18.6 \pm{1.5} $  &  $ 21.7 \pm{2.0} $  &   $  204 \pm{28} $ & $ 204 \pm{28} $  & $ 204 \pm{28} $ \\ 
 5.69   &   $ 0.81 \pm{0.01} $ &  $ 0.81 \pm{0.01} $  &  $ 0.81 \pm{0.01} $   &   $ 6.4 \pm{0.9} $ &   $ 7.4 \pm{0.8} $   &   $ 7.1 \pm{0.8} $   &   $ 15.6 \pm{1.7} $ &  $ 18.5 \pm{1.6} $  &  $ 21.6 \pm{1.9} $  &   $  214 \pm{29} $ & $ 214 \pm{29} $  & $ 213 \pm{29} $ \\ 
 5.89   &   $ 0.98 \pm{0.01} $ &  $ 1.01 \pm{0.01} $  &  $ 1.01 \pm{0.01} $   &   $ 7.0 \pm{1.3} $ &   $ 6.0 \pm{1.1} $   &   $ 6.3 \pm{1.4} $   &   $ 14.9 \pm{2.1} $ &  $ 18.4 \pm{2.7} $  &  $ 21.3 \pm{3.4} $  &   $  188 \pm{26} $ & $ 186 \pm{26} $  & $ 185 \pm{25} $ \\ 
 6.10   &   $ 1.12 \pm{0.01} $ &  $ 1.10 \pm{0.01} $  &  $ 1.02 \pm{0.02} $   &   $ 6.9 \pm{1.4} $ &   $ 6.8 \pm{1.0} $   &   $ 9.7 \pm{4.4} $   &   $ 15.8 \pm{16.} $ &  $ 18.1 \pm{2.3} $  &  $ 19.9 \pm{6.8} $  &   $  173 \pm{24} $ & $ 175 \pm{24} $  & $ 184 \pm{25} $ \\ 
 6.19   &   $ 1.06 \pm{0.01} $ &  $ 1.07 \pm{0.01} $  &  $ 1.08 \pm{0.01} $   &   $ 7.4 \pm{0.8} $ &   $ 7.7 \pm{0.6} $   &   $ 6.5 \pm{0.5} $   &   $ 16.2 \pm{1.3} $ &  $ 19.1 \pm{1.2} $  &  $ 22.3 \pm{1.5} $  &   $  179 \pm{24} $ & $ 178 \pm{24} $  & $ 177 \pm{24} $ \\ 
 6.30   &   $ 1.12 \pm{0.01} $ &  $ 1.13 \pm{0.01} $  &  $ 1.13 \pm{0.01} $   &   $ 8.6 \pm{1.4} $ &   $ 6.4 \pm{0.5} $   &   $ 6.5 \pm{0.5} $   &   $ 15.1 \pm{1.7} $ &  $ 19.3 \pm{1.2} $  &  $ 21.9 \pm{1.4} $  &   $  172 \pm{24} $ & $ 171 \pm{24} $  & $ 172 \pm{24} $ \\ 
 6.50   &   $ 1.13 \pm{0.01} $ &  $ 1.12 \pm{0.01} $  &  $ 1.11 \pm{0.01} $   &   $ 6.3 \pm{0.6} $ &   $ 7.0 \pm{0.5} $   &   $ 5.5 \pm{0.9} $   &   $ 15.6 \pm{1.3} $ &  $ 19.2 \pm{1.2} $  &  $ 22.2 \pm{2.9} $  &   $  172 \pm{24} $ & $ 172 \pm{24} $  & $ 174 \pm{24} $ \\ 
 6.60   &   $ 1.13 \pm{0.01} $ &  $ 1.14 \pm{0.01} $  &  $ 1.14 \pm{0.01} $   &   $ 5.8 \pm{0.5} $ &   $ 7.2 \pm{0.7} $   &   $ 6.3 \pm{0.5} $   &   $ 16.3 \pm{1.2} $ &  $ 18.7 \pm{1.4} $  &  $ 21.4 \pm{1.5} $  &   $  171 \pm{24} $ & $ 171 \pm{23} $  & $ 170 \pm{23} $ \\ 
 6.80   &   $ 1.29 \pm{0.01} $ &  $ 1.30 \pm{0.01} $  &  $ 1.30 \pm{0.01} $   &   $ 8.7 \pm{1.5} $ &   $ 9.8 \pm{1.2} $   &   $ 8.9 \pm{1.2} $   &   $ 15.2 \pm{2.3} $ &  $ 18.7 \pm{1.9} $  &  $ 21.0 \pm{2.3} $  &   $  156 \pm{22} $ & $ 156 \pm{21} $  & $ 156 \pm{21} $ \\ 
 7.10   &   $ 1.08 \pm{0.01} $ &  $ 1.09 \pm{0.01} $  &  $ 1.08 \pm{0.01} $   &   $ 6.6 \pm{0.9} $ &   $ 6.5 \pm{0.8} $   &   $ 6.5 \pm{0.9} $   &   $ 16.1 \pm{1.6} $ &  $ 18.7 \pm{1.8} $  &  $ 20.7 \pm{2.3} $  &   $  176 \pm{24} $ & $ 175 \pm{24} $  & $ 176 \pm{24} $ \\ 
 7.19   &   $ 1.16 \pm{0.01} $ &  $ 1.16 \pm{0.01} $  &  $ 1.16 \pm{0.01} $   &   $ 6.4 \pm{0.5} $ &   $ 7.0 \pm{0.4} $   &   $ 6.8 \pm{0.4} $   &   $ 15.9 \pm{1.0} $ &  $ 19.1 \pm{1.0} $  &  $ 21.8 \pm{1.2} $  &   $  168 \pm{23} $ & $ 169 \pm{23} $  & $ 169 \pm{23} $ \\ 
 7.39   &   $ 1.14 \pm{0.01} $ &  $ 1.14 \pm{0.01} $  &  $ 1.11 \pm{0.01} $   &   $ 6.9 \pm{0.7} $ &   $ 6.1 \pm{0.5} $   &   $ 6.6 \pm{0.9} $   &   $ 15.3 \pm{1.2} $ &  $ 19.2 \pm{1.2} $  &  $ 20.8 \pm{2.3} $  &   $  170 \pm{23} $ & $ 170 \pm{23} $  & $ 174 \pm{24} $ \\ 
 7.50   &   $ 1.20 \pm{0.01} $ &  $ 1.21 \pm{0.01} $  &  $ 1.18 \pm{0.01} $   &   $ 5.1 \pm{0.4} $ &   $ 5.1 \pm{0.5} $   &   $ 4.6 \pm{0.4} $   &   $ 16.1 \pm{1.1} $ &  $ 19.6 \pm{1.4} $  &  $ 22.6 \pm{1.5} $  &   $  165 \pm{23} $ & $ 164 \pm{22} $  & $ 166 \pm{23} $ \\ 
 7.60   &   $ 1.10 \pm{0.01} $ &  $ 1.10 \pm{0.01} $  &  $ 1.10 \pm{0.01} $   &   $ 4.4 \pm{0.3} $ &   $ 4.8 \pm{0.4} $   &   $ 4.5 \pm{0.3} $   &   $ 15.8 \pm{1.1} $ &  $ 19.4 \pm{1.3} $  &  $ 22.1 \pm{1.4} $  &   $  175 \pm{24} $ & $ 174 \pm{24} $  & $ 174 \pm{24} $ \\ 
 7.80   &   $ 1.17 \pm{0.01} $ &  $ 1.16 \pm{0.01} $  &  $ 1.17 \pm{0.01} $   &   $ 6.5 \pm{0.7} $ &   $ 6.1 \pm{0.5} $   &   $ 5.0 \pm{0.5} $   &   $ 15.4 \pm{1.4} $ &  $ 18.6 \pm{1.3} $  &  $ 21.4 \pm{2.0} $  &   $  168 \pm{23} $ & $ 168 \pm{23} $  & $ 168 \pm{23} $ \\ 
 8.00   &   $ 1.33 \pm{0.01} $ &  $ 1.32 \pm{0.01} $  &  $ 1.23 \pm{0.01} $   &   $ 6.0 \pm{0.9} $ &   $ 5.7 \pm{0.7} $   &   $ 13. \pm{1.9} $   &   $ 15.5 \pm{2.0} $ &  $ 18.7 \pm{1.9} $  &  $ 19.5 \pm{3.0} $  &   $  154 \pm{21} $ & $ 154 \pm{21} $  & $ 162 \pm{22} $ \\ 
 8.19   &   $ 1.34 \pm{0.01} $ &  $ 1.33 \pm{0.01} $  &  $ 1.33 \pm{0.01} $   &   $ 7.2 \pm{0.7} $ &   $ 7.2 \pm{0.5} $   &   $ 6.2 \pm{0.4} $   &   $ 15.2 \pm{1.2} $ &  $ 18.7 \pm{1.0} $  &  $ 21.2 \pm{1.2} $  &   $  153 \pm{21} $ & $ 154 \pm{21} $  & $ 154 \pm{21} $ \\ 
 8.30   &   $ 1.20 \pm{0.01} $ &  $ 1.20 \pm{0.01} $  &  $ 1.16 \pm{0.01} $   &   $ 5.1 \pm{0.5} $ &   $ 6.1 \pm{0.4} $   &   $ 6.6 \pm{0.6} $   &   $ 15.6 \pm{1.2} $ &  $ 19.0 \pm{1.0} $  &  $ 21.1 \pm{1.7} $  &   $  165 \pm{23} $ & $ 165 \pm{23} $  & $ 168 \pm{23} $ \\ 
 8.39   &   $ 1.21 \pm{0.01} $ &  $ 1.20 \pm{0.01} $  &  $ 1.18 \pm{0.01} $   &   $ 6.5 \pm{0.6} $ &   $ 6.1 \pm{0.5} $   &   $ 8.5 \pm{1.2} $   &   $ 15.4 \pm{1.2} $ &  $ 18.8 \pm{1.2} $  &  $ 19.3 \pm{2.2} $  &   $  164 \pm{23} $ & $ 164 \pm{22} $  & $ 167 \pm{23} $ \\ 
 8.60   &   $ 1.18 \pm{0.01} $ &  $ 1.20 \pm{0.01} $  &  $ 1.20 \pm{0.01} $   &   $ 7.6 \pm{1.1} $ &   $ 7.6 \pm{0.7} $   &   $ 6.9 \pm{0.7} $   &   $ 13.6 \pm{1.5} $ &  $ 18.5 \pm{1.3} $  &  $ 21.4 \pm{1.7} $  &   $  166 \pm{23} $ & $ 164 \pm{22} $  & $ 164 \pm{23} $ \\ 
 8.69   &   $ 1.24 \pm{0.01} $ &  $ 1.24 \pm{0.01} $  &  $ 1.24 \pm{0.01} $   &   $ 6.3 \pm{0.6} $ &   $ 6.0 \pm{0.5} $   &   $ 5.6 \pm{0.4} $   &   $ 15.6 \pm{1.1} $ &  $ 19.1 \pm{1.3} $  &  $ 21.6 \pm{1.4} $  &   $  161 \pm{22} $ & $ 161 \pm{22} $  & $ 161 \pm{22} $ \\ 
 8.80   &   $ 1.23 \pm{0.01} $ &  $ 1.24 \pm{0.01} $  &  $ 1.23 \pm{0.01} $   &   $ 8.7 \pm{1.5} $ &   $ 7.5 \pm{1.0} $   &   $ 7.3 \pm{0.9} $   &   $ 15.1 \pm{2.1} $ &  $ 18.4 \pm{2.0} $  &  $ 20.9 \pm{2.2} $  &   $  162 \pm{22} $ & $ 161 \pm{22} $  & $ 162 \pm{22} $ \\ 
 9.00   &   $ 1.24 \pm{0.01} $ &  $ 1.24 \pm{0.01} $  &  $ 1.23 \pm{0.01} $   &   $ 7.3 \pm{1.3} $ &   $ 7.6 \pm{0.3} $   &   $ 10. \pm{1.2} $   &   $ 14.5 \pm{2.0} $ &  $ 19.6 \pm{1.3} $  &  $ 19.8 \pm{1.3} $  &   $  161 \pm{22} $ & $ 161 \pm{22} $  & $ 162 \pm{22} $ \\ 
 9.10   &   $ 1.31 \pm{0.01} $ &  $ 1.32 \pm{0.01} $  &  $ 1.31 \pm{0.01} $   &   $ 6.8 \pm{0.6} $ &   $ 7.0 \pm{0.5} $   &   $ 6.6 \pm{0.4} $   &   $ 15.1 \pm{1.1} $ &  $ 18.5 \pm{1.0} $  &  $ 20.9 \pm{1.1} $  &   $  155 \pm{21} $ & $ 156 \pm{21} $  & $ 155 \pm{21} $ \\ 
 9.30   &   $ 1.34 \pm{0.01} $ &  $ 1.34 \pm{0.01} $  &  $ 1.33 \pm{0.01} $   &   $ 6.2 \pm{0.5} $ &   $ 7.1 \pm{0.7} $   &   $ 7.3 \pm{0.7} $   &   $ 15.5 \pm{1.1} $ &  $ 18.6 \pm{1.2} $  &  $ 20.7 \pm{1.6} $  &   $  153 \pm{21} $ & $ 153 \pm{21} $  & $ 154 \pm{21} $ \\ 
 9.39   &   $ 1.24 \pm{0.01} $ &  $ 1.25 \pm{0.01} $  &  $ 1.29 \pm{0.01} $   &   $ 4.5 \pm{0.4} $ &   $ 4.9 \pm{0.3} $   &   $ 6.5 \pm{0.8} $   &   $ 16.0 \pm{1.3} $ &  $ 19.6 \pm{1.1} $  &  $ 20.7 \pm{2.1} $  &   $  161 \pm{22} $ & $ 160 \pm{22} $  & $ 157 \pm{21} $ \\ 
 9.60   &   $ 1.17 \pm{0.01} $ &  $ 1.17 \pm{0.01} $  &  $ 1.17 \pm{0.01} $   &   $ 7.6 \pm{0.8} $ &   $ 7.9 \pm{0.8} $   &   $ 6.5 \pm{0.6} $   &   $ 14.4 \pm{1.2} $ &  $ 17.7 \pm{1.4} $  &  $ 20.2 \pm{1.5} $  &   $  167 \pm{23} $ & $ 167 \pm{23} $  & $ 167 \pm{23} $ \\ 
 9.69   &   $ 1.24 \pm{0.01} $ &  $ 1.24 \pm{0.01} $  &  $ 1.23 \pm{0.01} $   &   $ 4.0 \pm{0.3} $ &   $ 4.2 \pm{0.3} $   &   $ 4.1 \pm{0.3} $   &   $ 15.8 \pm{1.0} $ &  $ 18.6 \pm{1.1} $  &  $ 21.3 \pm{1.2} $  &   $  161 \pm{22} $ & $ 162 \pm{22} $  & $ 161 \pm{22} $ \\ 
 9.80   &   $ 1.36 \pm{0.01} $ &  $ 1.34 \pm{0.01} $  &  $ 1.34 \pm{0.01} $   &   $ 5.5 \pm{1.5} $ &   $ 8.4 \pm{0.4} $   &   $ 5.9 \pm{0.6} $   &   $ 13.9 \pm{2.7} $ &  $ 18.5 \pm{2.0} $  &  $ 20.7 \pm{1.9} $  &   $  151 \pm{21} $ & $ 153 \pm{21} $  & $ 153 \pm{21} $ \\ 
10.00   &   $ 1.32 \pm{0.01} $ &  $ 1.32 \pm{0.01} $  &  $ 1.33 \pm{0.01} $   &   $ 7.0 \pm{0.7} $ &   $ 5.8 \pm{1.2} $   &   $ 11. \pm{4.0} $   &   $ 15.3 \pm{2.1} $ &  $ 18.5 \pm{2.9} $  &  $ 17.5 \pm{4.4} $  &   $  155 \pm{21} $ & $ 154 \pm{21} $  & $ 154 \pm{21} $ \\ 
10.10   &   $ 1.44 \pm{0.01} $ &  $ 1.44 \pm{0.01} $  &  $ 1.50 \pm{0.01} $   &   $ 5.3 \pm{0.6} $ &   $ 6.1 \pm{0.5} $   &   $ 7.1 \pm{0.8} $   &   $ 15.4 \pm{1.3} $ &  $ 18.6 \pm{1.2} $  &  $ 19.9 \pm{1.6} $  &   $  146 \pm{20} $ & $ 145 \pm{20} $  & $ 142 \pm{20} $ \\ 
10.19   &   $ 1.31 \pm{0.01} $ &  $ 1.32 \pm{0.01} $  &  $ 1.32 \pm{0.01} $   &   $ 6.6 \pm{0.5} $ &   $ 6.4 \pm{0.4} $   &   $ 6.3 \pm{0.4} $   &   $ 15.4 \pm{1.0} $ &  $ 18.6 \pm{1.1} $  &  $ 20.7 \pm{1.1} $  &   $  155 \pm{21} $ & $ 154 \pm{21} $  & $ 155 \pm{21} $ \\ 
10.50   &   $ 1.27 \pm{0.01} $ &  $ 1.28 \pm{0.01} $  &  $ 1.28 \pm{0.01} $   &   $ 7.2 \pm{0.9} $ &   $ 7.2 \pm{0.7} $   &   $ 8.4 \pm{0.3} $   &   $ 14.6 \pm{1.5} $ &  $ 18.1 \pm{1.5} $  &  $ 21.1 \pm{1.0} $  &   $  158 \pm{22} $ & $ 158 \pm{22} $  & $ 158 \pm{22} $ \\ 
10.60   &   $ 1.29 \pm{0.01} $ &  $ 1.30 \pm{0.01} $  &  $ 1.30 \pm{0.01} $   &   $ 7.3 \pm{1.1} $ &   $ 6.6 \pm{0.5} $   &   $ 6.0 \pm{0.5} $   &   $ 13.5 \pm{1.4} $ &  $ 18.1 \pm{1.2} $  &  $ 20.4 \pm{1.4} $  &   $  157 \pm{22} $ & $ 156 \pm{21} $  & $ 156 \pm{21} $ \\ 
10.69   &   $ 1.30 \pm{0.01} $ &  $ 1.30 \pm{0.01} $  &  $ 1.30 \pm{0.00} $   &   $ 6.6 \pm{0.7} $ &   $ 7.8 \pm{0.8} $   &   $ 7.0 \pm{0.7} $   &   $ 14.9 \pm{1.3} $ &  $ 17.8 \pm{1.5} $  &  $ 20.2 \pm{1.7} $  &   $  156 \pm{21} $ & $ 156 \pm{21} $  & $ 156 \pm{21} $ \\ 
11.00   &   $ 1.24 \pm{0.01} $ &  $ 1.23 \pm{0.01} $  &  $ 1.26 \pm{0.01} $   &   $ 6.5 \pm{0.4} $ &   $ 6.6 \pm{0.7} $   &   $ 7.1 \pm{1.7} $   &   $ 15.4 \pm{1.2} $ &  $ 18.5 \pm{1.7} $  &  $ 21.2 \pm{3.9} $  &   $  161 \pm{22} $ & $ 162 \pm{22} $  & $ 160 \pm{22} $ \\ 
11.10   &   $ 1.26 \pm{0.01} $ &  $ 1.26 \pm{0.01} $  &  $ 1.26 \pm{0.01} $   &   $ 5.0 \pm{0.4} $ &   $ 5.6 \pm{0.3} $   &   $ 5.4 \pm{0.3} $   &   $ 15.3 \pm{1.0} $ &  $ 18.6 \pm{1.0} $  &  $ 21.0 \pm{1.1} $  &   $  160 \pm{22} $ & $ 160 \pm{22} $  & $ 160 \pm{22} $ \\ 
11.30   &   $ 1.18 \pm{0.01} $ &  $ 1.18 \pm{0.01} $  &  $ 1.21 \pm{0.01} $   &   $ 6.0 \pm{0.9} $ &   $ 6.7 \pm{0.5} $   &   $ 7.5 \pm{0.7} $   &   $ 14.1 \pm{1.5} $ &  $ 17.6 \pm{1.1} $  &  $ 18.2 \pm{1.4} $  &   $  166 \pm{23} $ & $ 167 \pm{23} $  & $ 164 \pm{22} $ \\ 
11.39   &   $ 1.12 \pm{0.01} $ &  $ 1.13 \pm{0.01} $  &  $ 1.14 \pm{0.01} $   &   $ 4.7 \pm{0.5} $ &   $ 6.0 \pm{0.5} $   &   $ 4.9 \pm{0.5} $   &   $ 15.3 \pm{1.2} $ &  $ 18.0 \pm{1.3} $  &  $ 20.4 \pm{1.9} $  &   $  172 \pm{24} $ & $ 171 \pm{24} $  & $ 170 \pm{23} $ \\ 
11.50   &   $ 1.18 \pm{0.01} $ &  $ 1.18 \pm{0.01} $  &  $ 1.17 \pm{0.01} $   &   $ 5.8 \pm{0.7} $ &   $ 6.4 \pm{0.6} $   &   $ 6.6 \pm{0.5} $   &   $ 14.8 \pm{1.4} $ &  $ 17.5 \pm{1.4} $  &  $ 19.8 \pm{1.4} $  &   $  166 \pm{23} $ & $ 167 \pm{23} $  & $ 167 \pm{23} $ \\ 
11.69   &   $ 1.09 \pm{0.01} $ &  $ 1.09 \pm{0.01} $  &  $ 1.09 \pm{0.01} $   &   $ 4.6 \pm{0.5} $ &   $ 6.0 \pm{0.5} $   &   $ 6.0 \pm{0.5} $   &   $ 15.5 \pm{1.4} $ &  $ 17.9 \pm{1.3} $  &  $ 20.3 \pm{1.4} $  &   $  176 \pm{24} $ & $ 176 \pm{24} $  & $ 176 \pm{24} $ \\ 
11.80   &   $ 1.07 \pm{0.01} $ &  $ 1.09 \pm{0.01} $  &  $ 1.09 \pm{0.01} $   &   $ 5.9 \pm{1.0} $ &   $ 5.8 \pm{1.0} $   &   $ 5.4 \pm{0.7} $   &   $ 14.9 \pm{2.0} $ &  $ 17.1 \pm{2.3} $  &  $ 20.8 \pm{2.5} $  &   $  178 \pm{24} $ & $ 176 \pm{24} $  & $ 176 \pm{24} $ \\ 
12.10   &   $ 1.12 \pm{0.01} $ &  $ 1.12 \pm{0.01} $  &  $ 1.10 \pm{0.01} $   &   $ 6.0 \pm{0.9} $ &   $ 6.1 \pm{0.4} $   &   $ 6.5 \pm{0.6} $   &   $ 14.5 \pm{1.6} $ &  $ 17.7 \pm{1.1} $  &  $ 19.9 \pm{1.4} $  &   $  173 \pm{24} $ & $ 172 \pm{24} $  & $ 174 \pm{24} $ \\ 
12.19   &   $ 1.11 \pm{0.01} $ &  $ 1.10 \pm{0.01} $  &  $ 1.11 \pm{0.01} $   &   $ 5.9 \pm{0.6} $ &   $ 6.1 \pm{0.5} $   &   $ 6.0 \pm{0.5} $   &   $ 14.7 \pm{1.2} $ &  $ 17.6 \pm{1.2} $  &  $ 19.9 \pm{1.2} $  &   $  173 \pm{24} $ & $ 174 \pm{24} $  & $ 174 \pm{24} $ \\ 
12.30   &   $ 1.24 \pm2{0.0} $ &  $ 1.26 \pm{0.01} $  &  $ 1.31 \pm{0.01} $   &   $ 4.1 \pm{0.4} $ &   $ 4.7 \pm{0.3} $   &   $ 7.0 \pm{1.3} $   &   $ 15.6 \pm{1.2} $ &  $ 17.8 \pm{1.0} $  &  $ 18.6 \pm{2.9} $  &   $  161 \pm{22} $ & $ 160 \pm{22} $  & $ 155 \pm{21} $ \\ 
12.50   &   $ 1.50 \pm{40.0} $ &  $ 1.50 \pm{0.01} $  &  $ 1.48 \pm{0.01} $   &   $ 4.5 \pm{0.4} $ &   $ 5.4 \pm{0.4} $   &   $ 5.5 \pm{0.2} $   &   $ 14.1 \pm{1.0} $ &  $ 17.7 \pm{1.0} $  &  $ 20.9 \pm{0.8} $  &   $  142 \pm{20} $ & $ 142 \pm{20} $  & $ 143 \pm{20} $ \\ 
12.60   &   $ 1.46 \pm{0.01} $ &  $ 1.47 \pm{0.01} $  &  $ 1.47 \pm{0.01} $   &   $ 5.5 \pm{0.7} $ &   $ 6.2 \pm{0.5} $   &   $ 6.0 \pm{0.4} $   &   $ 14.7 \pm{1.4} $ &  $ 18.0 \pm{1.1} $  &  $ 20.1 \pm{1.2} $  &   $  145 \pm{20} $ & $ 144 \pm{20} $  & $ 144 \pm{20} $ \\ 
12.69   &   $ 1.37 \pm{0.01} $ &  $ 1.38 \pm{0.01} $  &  $ 1.38 \pm{0.01} $   &   $ 8.0 \pm{1.1} $ &   $ 8.0 \pm{0.8} $   &   $ 7.2 \pm{0.8} $   &   $ 14.3 \pm{1.4} $ &  $ 17.3 \pm{1.5} $  &  $ 19.5 \pm{1.6} $  &   $  150 \pm{21} $ & $ 150 \pm{20} $  & $ 150 \pm{21} $ \\ 
13.00   &   $ 1.45 \pm{0.01} $ &  $ 1.46 \pm{0.01} $  &  $ 1.47 \pm{0.01} $   &   $ 7.7 \pm{1.2} $ &   $ 8.1 \pm{0.8} $   &   $ 8.0 \pm{1.3} $   &   $ 13.3 \pm{1.5} $ &  $ 17.2 \pm{1.3} $  &  $ 18.4 \pm{2.4} $  &   $  145 \pm{20} $ & $ 144 \pm{20} $  & $ 144 \pm{20} $ \\ 
13.10   &   $ 1.47 \pm{0.01} $ &  $ 1.48 \pm{0.01} $  &  $ 1.47 \pm{0.01} $   &   $ 6.4 \pm{0.6} $ &   $ 6.8 \pm{0.4} $   &   $ 6.6 \pm{0.4} $   &   $ 14.0 \pm{1.0} $ &  $ 17.5 \pm{0.9} $  &  $ 19.3 \pm{1.0} $  &   $  144 \pm{20} $ & $ 143 \pm{20} $  & $ 144 \pm{20} $ \\ 
13.30   &   $ 1.42 \pm{0.01} $ &  $ 1.42 \pm{0.01} $  &  $ 1.43 \pm{0.01} $   &   $ 5.3 \pm{0.5} $ &   $ 7.4 \pm{0.5} $   &   $ 7.6 \pm{0.8} $   &   $ 14.5 \pm{1.2} $ &  $ 17.4 \pm{1.0} $  &  $ 19.1 \pm{1.6} $  &   $  147 \pm{20} $ & $ 148 \pm{20} $  & $ 147 \pm{20} $ \\ 
13.39   &   $ 1.37 \pm{0.01} $ &  $ 1.35 \pm{0.01} $  &  $ 1.39 \pm{0.01} $   &   $ 5.5 \pm{0.7} $ &   $ 5.7 \pm{0.4} $   &   $ 5.6 \pm{0.5} $   &   $ 14.3 \pm{1.3} $ &  $ 17.7 \pm{1.0} $  &  $ 19.5 \pm{1.3} $  &   $  151 \pm{21} $ & $ 152 \pm{21} $  & $ 149 \pm{20} $ \\ 
14.10   &   $ 1.31 \pm{0.01} $ &  $ 1.29 \pm{0.01} $  &  $ 1.29 \pm{0.01} $   &   $ 3.5 \pm{0.3} $ &   $ 4.3 \pm{0.2} $   &   $ 3.7 \pm{0.2} $   &   $ 15.3 \pm{1.1} $ &  $ 17.4 \pm{0.8} $  &  $ 20.5 \pm{0.9} $  &   $  151 \pm{21} $ & $ 157 \pm{22} $  & $ 157 \pm{22} $ \\ 
14.19   &   $ 1.28 \pm{0.01} $ &  $ 1.28 \pm{0.01} $  &  $ 1.25 \pm{0.01} $   &   $ 5.7 \pm{0.6} $ &   $ 5.8 \pm{0.5} $   &   $ 7.2 \pm{0.8} $   &   $ 14.7 \pm{1.3} $ &  $ 17.5 \pm{1.1} $  &  $ 19.1 \pm{1.6} $  &   $  157 \pm{22} $ & $ 158 \pm{22} $  & $ 161 \pm{22} $ \\ 
14.39   &   $ 1.40 \pm{0.01} $ &  $ 1.39 \pm{0.01} $  &  $ 1.40 \pm{0.01} $   &   $ 5.3 \pm{0.8} $ &   $ 6.5 \pm{0.5} $   &   $ 6.3 \pm{0.6} $   &   $ 14.4 \pm{1.4} $ &  $ 17.4 \pm{1.0} $  &  $ 19.0 \pm{1.3} $  &   $  149 \pm{20} $ & $ 149 \pm{20} $  & $ 149 \pm{20} $ \\ 
14.50   &   $ 1.50 \pm{0.01} $ &  $ 1.49 \pm{0.01} $  &  $ 1.49 \pm{0.01} $   &   $ 5.0 \pm{0.5} $ &   $ 5.9 \pm{0.4} $   &   $ 5.6 \pm{0.3} $   &   $ 14.7 \pm{1.1} $ &  $ 17.6 \pm{1.0} $  &  $ 19.8 \pm{1.1} $  &   $  142 \pm{20} $ & $ 142 \pm{20} $  & $ 142 \pm{20} $ \\ 
14.89   &   $ 1.39 \pm{0.01} $ &  $ 1.40 \pm{0.01} $  &  $ 1.47 \pm{0.03} $   &   $ 8.7 \pm{4.1} $ &   $ 7.5 \pm{1.5} $   &   $ 7.5 \pm{5.2} $   &   $ 12.0 \pm{4.3} $ &  $ 16.3 \pm{2.5} $  &  $ 15.4 \pm{6.8} $  &   $  149 \pm{20} $ & $ 148 \pm{20} $  & $ 144 \pm{20} $ \\ 
15.00   &   $ 1.30 \pm{0.00} $ &  $ 1.30 \pm{0.01} $  &  $ 1.30 \pm{0.01} $   &   $ 6.1 \pm{0.2} $ &   $ 5.8 \pm{0.4} $   &   $ 5.4 \pm{0.4} $   &   $ 15.3 \pm{0.7} $ &  $ 17.0 \pm{1.0} $  &  $ 19.2 \pm{1.2} $  &   $  156 \pm{21} $ & $ 156 \pm{21} $  & $ 156 \pm{21} $ \\ 
15.19   &   $ 1.31 \pm{0.01} $ &  $ 1.32 \pm{0.01} $  &  $ 1.30 \pm{0.01} $   &   $ 5.0 \pm{0.7} $ &   $ 5.6 \pm{0.4} $   &   $ 5.6 \pm{0.5} $   &   $ 14.3 \pm{1.3} $ &  $ 17.3 \pm{1.0} $  &  $ 18.8 \pm{1.3} $  &   $  155 \pm{21} $ & $ 155 \pm{21} $  & $ 156 \pm{21} $ \\ 
15.30   &   $ 1.39 \pm{0.00} $ &  $ 1.39 \pm{0.01} $  &  $ 1.38 \pm{0.01} $   &   $ 5.9 \pm{0.7} $ &   $ 7.2 \pm{0.6} $   &   $ 6.8 \pm{0.8} $   &   $ 14.0 \pm{1.3} $ &  $ 16.5 \pm{1.1} $  &  $ 18.3 \pm{1.7} $  &   $  149 \pm{20} $ & $ 150 \pm{20} $  & $ 150 \pm{21} $ \\ 
15.39   &   $ 1.39 \pm{0.01} $ &  $ 1.40 \pm{0.01} $  &  $ 1.40 \pm{0.01} $   &   $ 6.0 \pm{0.8} $ &   $ 6.6 \pm{0.5} $   &   $ 5.8 \pm{0.4} $   &   $ 13.6 \pm{1.3} $ &  $ 16.7 \pm{1.0} $  &  $ 21.5 \pm{1.0} $  &   $  149 \pm{20} $ & $ 149 \pm{20} $  & $ 149 \pm{20} $ \\ 
15.60   &   $ 1.30 \pm{0.01} $ &  $ 1.31 \pm{0.01} $  &  $ 1.32 \pm{0.01} $   &   $ 6.9 \pm{0.9} $ &   $ 6.2 \pm{0.5} $   &   $ 5.6 \pm{0.4} $   &   $ 13.5 \pm{1.3} $ &  $ 17.0 \pm{1.1} $  &  $ 19.5 \pm{1.1} $  &   $  156 \pm{22} $ & $ 155 \pm{21} $  & $ 154 \pm{21} $ \\ 
15.69   &   $ 1.37 \pm{0.01} $ &  $ 1.37 \pm{0.01} $  &  $ 1.36 \pm{0.01} $   &   $ 6.2 \pm{1.6} $ &   $ 7.6 \pm{1.0} $   &   $ 6.1 \pm{0.8} $   &   $ 13.4 \pm{2.6} $ &  $ 15.8 \pm{1.7} $  &  $ 17.6 \pm{1.9} $  &   $  151 \pm{21} $ & $ 151 \pm{21} $  & $ 152 \pm{21} $ \\ 
16.00   &   $ 1.40 \pm{0.01} $ &  $ 1.40 \pm{0.01} $  &  $ 1.40 \pm{0.01} $   &   $ 5.3 \pm{0.8} $ &   $ 5.9 \pm{0.4} $   &   $ 5.6 \pm{0.4} $   &   $ 13.8 \pm{1.5} $ &  $ 16.9 \pm{0.9} $  &  $ 18.8 \pm{1.0} $  &   $  149 \pm{20} $ & $ 148 \pm{20} $  & $ 149 \pm{20} $ \\ 
16.19   &   $ 1.37 \pm{0.01} $ &  $ 1.39 \pm{0.01} $  &  $ 1.39 \pm{0.01} $   &   $ 7.5 \pm{1.2} $ &   $ 7.3 \pm{0.6} $   &   $ 7.5 \pm{0.8} $   &   $ 13.2 \pm{1.5} $ &  $ 16.4 \pm{1.1} $  &  $ 17.9 \pm{1.5} $  &   $  151 \pm{21} $ & $ 150 \pm{20} $  & $ 149 \pm{20} $ \\ 
16.30   &   $ 1.25 \pm{0.01} $ &  $ 1.27 \pm{0.01} $  &  $ 1.25 \pm{0.01} $   &   $ 6.1 \pm{0.9} $ &   $ 6.8 \pm{0.7} $   &   $ 6.6 \pm{1.0} $   &   $ 13.8 \pm{1.5} $ &  $ 16.2 \pm{1.2} $  &  $ 18.3 \pm{2.0} $  &   $  161 \pm{22} $ & $ 159 \pm{22} $  & $ 160 \pm{22} $ \\ 
16.39   &   $ 1.23 \pm{0.01} $ &  $ 1.25 \pm{0.01} $  &  $ 1.24 \pm{0.01} $   &   $ 5.0 \pm{0.8} $ &   $ 5.2 \pm{0.4} $   &   $ 4.8 \pm{0.3} $   &   $ 14.0 \pm{1.5} $ &  $ 17.0 \pm{1.0} $  &  $ 19.4 \pm{1.2} $  &   $  162 \pm{22} $ & $ 160 \pm{22} $  & $ 161 \pm{22} $ \\ 
16.60   &   $ 1.12 \pm{0.01} $ &  $ 1.12 \pm{0.01} $  &  $ 1.12 \pm{0.01} $   &   $ 7.6 \pm{1.3} $ &   $ 6.7 \pm{0.6} $   &   $ 5.7 \pm{0.5} $   &   $ 13.8 \pm{1.9} $ &  $ 20.4 \pm{1.2} $  &  $ 18.8 \pm{1.3} $  &   $  173 \pm{24} $ & $ 172 \pm{24} $  & $ 172 \pm{24} $ \\ 
17.00   &   $ 1.30 \pm{0.01} $ &  $ 1.26 \pm{0.01} $  &  $ 1.31 \pm{0.01} $   &   $ 5.6 \pm{0.2} $ &   $ 8.2 \pm{0.2} $   &   $ 5.6 \pm{0.4} $   &   $ 15.2 \pm{0.8} $ &  $ 17.4 \pm{3.7} $  &  $ 18.7 \pm{1.1} $  &   $  156 \pm{21} $ & $ 159 \pm{22} $  & $ 155 \pm{21} $ \\ 
17.10   &   $ 1.45 \pm{0.01} $ &  $ 1.46 \pm{0.01} $  &  $ 1.46 \pm{0.01} $   &   $ 7.0 \pm{0.9} $ &   $ 7.9 \pm{0.7} $   &   $ 9.3 \pm{0.5} $   &   $ 13.1 \pm{1.3} $ &  $ 15.8 \pm{1.2} $  &  $ 18.4 \pm{1.8} $  &   $  145 \pm{20} $ & $ 145 \pm{20} $  & $ 145 \pm{20} $ \\ 
17.30   &   $ 1.50 \pm{0.01} $ &  $ 1.51 \pm{0.01} $  &  $ 1.52 \pm{0.01} $   &   $ 5.8 \pm{0.7} $ &   $ 7.3 \pm{0.5} $   &   $ 7.0 \pm{0.6} $   &   $ 13.3 \pm{1.3} $ &  $ 16.7 \pm{1.0} $  &  $ 18.3 \pm{1.4} $  &   $  142 \pm{19} $ & $ 141 \pm{19} $  & $ 141 \pm{19} $ \\ 
17.39   &   $ 1.42 \pm{0.01} $ &  $ 1.43 \pm{0.01} $  &  $ 1.43 \pm{0.01} $   &   $ 6.9 \pm{1.1} $ &   $ 6.6 \pm{0.4} $   &   $ 5.8 \pm{0.4} $   &   $ 13.2 \pm{1.4} $ &  $ 16.4 \pm{1.0} $  &  $ 18.4 \pm{1.0} $  &   $  147 \pm{20} $ & $ 146 \pm{20} $  & $ 147 \pm{20} $ \\ 
17.60   &   $ 1.46 \pm{0.01} $ &  $ 1.43 \pm{0.01} $  &  $ 1.44 \pm{0.01} $   &   $ 6.4 \pm{1.1} $ &   $ 6.8 \pm{0.5} $   &   $ 6.7 \pm{0.5} $   &   $ 12.9 \pm{1.6} $ &  $ 16.0 \pm{1.0} $  &  $ 17.3 \pm{1.1} $  &   $  145 \pm{20} $ & $ 146 \pm{20} $  & $ 146 \pm{20} $ \\ 
17.80   &   $ 1.70 \pm{0.03} $ &  $ 1.80 \pm{0.01} $  &  $ 1.79 \pm{0.01} $   &   $ 5.6 \pm{1.8} $ &   $ 4.0 \pm{0.4} $   &   $ 8.3 \pm{2.4} $   &   $ 13.1 \pm{3.2} $ &  $ 17.3 \pm{1.5} $  &  $ 15.0 \pm{3.1} $  &   $  131 \pm{18} $ & $ 126 \pm{17} $  & $ 126 \pm{17} $ \\ 
18.00   &   $ 1.86 \pm{0.01} $ &  $ 1.84 \pm{0.01} $  &  $ 1.85 \pm{0.01} $   &   $ 6.9 \pm{0.8} $ &   $ 6.9 \pm{0.3} $   &   $ 6.6 \pm{0.3} $   &   $ 12.5 \pm{1.2} $ &  $ 16.6 \pm{0.7} $  &  $ 18.3 \pm{0.8} $  &   $  123 \pm{17} $ & $ 124 \pm{17} $  & $ 123 \pm{17} $ \\ 
\end{longtable}
 \leftline{{\large In column 1, we have listed the MJD-60180 (to save space) of the exposure IDs we used.}}
 \leftline{{\large Columns 2, 3, \& 4 represent the QPO frequency in LE, ME, and HE energy bands respectively.}}
 \leftline{{\large Columns 5, 6, \& 7 represent the Q-values of QPOs in LE, ME, and HE energy bands respectively.}}
 \leftline{{\large Columns 8, 9, \& 10 represent the QPO RMS (\%) in LE, ME, and HE energy bands respectively.}}
 \leftline{{\large Columns 11, 12, \& 13 represent the shock location in LE, ME, and HE energy bands respectively.}}

\newpage
\begin{sidewaystable}[htbp]
\scriptsize
 \addtolength{\tabcolsep}{2.0pt}
 \centering
 \caption{Properties from spectral analysis}
 \label{tab:table4}
 \resizebox{1 \textwidth}{!}{
 \begin{tabular}{|c|c|c|c|c|c|c|c|c|c|c|c|c|c|c|c|c|c|}
 \hline
 Time   &        TBabs                 &            \multicolumn{3}{|c|}{gabs}           &      \multicolumn{2}{|c|}{diskbb}      &     \multicolumn{2}{|c|}{power-law}      &       \multicolumn{3}{|c|}{Gaussian}         &                          \multicolumn{5}{|c|}{pexrav}                               &                   \\
\hline 
 MJD    &        $N_H$                 &    E$_{\rm abs}$      &    $\sigma_{\rm abs}$    &    N$_{\rm abs}$    &     $T_{in}$      &       Norm         &      $\Gamma$        &     Norm          &   $E_g$     &       $\sigma_g$       &    Norm    &    $\Gamma$     &     $E_{\rm cut}$      &    R$_{\rm ref}$    &    $\cos(i)$     &    Norm       &     $\chi^2/DOF$  \\
 (Day)  & ($\times 10^{22}$~cm$^{-2}$) &       (keV)       &       (keV)      &      (keV)       &         (keV)         & $(R_{in}/D_{10})^2 cos(i)$ &                   & (photons/keV/cm$^2$/s) &    (keV)     &        (keV)       &     ($ph/cm^2/s$)   &             &      (keV)      &                 &       & (photons/keV/cm$^2$/s) &              \\ 
	&                              &                   &                  &                  &                       &      ($\times 10^4$)       &                   &                        &              &                    &                     &             &                 &                 &       &                        &              \\  
\hline 
 (1)    &          (2)                 &       (3)       &       (4)      &      (5)       &       (6)         &        (7)         &         (8)          &      (9)          &    (10)     &        (11)       &     (12)   &       (13)      &      (14)      &       (15)          & (16)      &    (17)       &          (18)     \\ 
\hline 
60181.4 & $ 0.19 \pm{0.03} $ & $1.78 \pm{0.16}$ & $0.41 \pm{0.03}$ & $0.18 \pm{0.02}$  & $0.19 \pm{0.02}$ & $ 18 \pm{ 2}$  & $2.12 \pm{0.15}$ & $13.1 \pm{1.3}$ & $5.25 \pm{0.66}$ & $1.92 \pm{0.15}$ & $0.82 \pm{0.12}$ & $1.14 \pm{0.11}$ & $26.2 \pm{1.1}$ &$  0.40 \pm{0.05}$  &  $0.49 \pm{0.04}$ &  $ 3.3 \pm{0.4}$ & $1384.5/1401$  \\
60181.6 & $ 0.19 \pm{0.06} $ & $1.80 \pm{0.30}$ & $0.23 \pm{0.04}$ & $0.01 \pm{0.01}$  & $0.21 \pm{0.02}$ & $ 19 \pm{ 1}$  & $2.16 \pm{0.15}$ & $18.2 \pm{1.2}$ & $5.48 \pm{0.62}$ & $1.62 \pm{0.13}$ & $0.63 \pm{0.05}$ & $0.94 \pm{0.11}$ & $20.3 \pm{1.2}$ &$  0.50 \pm{0.06}$  &  $0.19 \pm{0.02}$ &  $ 2.7 \pm{0.8}$ & $1375.4/1401$  \\
60181.7 & $ 0.19 \pm{0.06} $ & $1.65 \pm{0.24}$ & $0.43 \pm{0.03}$ & $0.01 \pm{0.01}$  & $0.31 \pm{0.05}$ & $ 11 \pm{ 3}$  & $2.10 \pm{0.14}$ & $15.1 \pm{1.0}$ & $5.62 \pm{0.55}$ & $0.92 \pm{0.13}$ & $0.21 \pm{0.03}$ & $1.01 \pm{0.08}$ & $20.1 \pm{0.9}$ &$  0.52 \pm{0.09}$  &  $0.05 \pm{0.01}$ &  $ 4.4 \pm{0.8}$ & $1341.3/1283$  \\
60182.1 & $ 0.19 \pm{0.08} $ & $1.65 \pm{0.15}$ & $0.46 \pm{0.03}$ & $0.01 \pm{0.02}$  & $0.26 \pm{0.02}$ & $ 11 \pm{ 4}$  & $2.16 \pm{0.14}$ & $18.5 \pm{0.6}$ & $5.55 \pm{1.05}$ & $0.88 \pm{0.26}$ & $0.26 \pm{0.07}$ & $1.01 \pm{0.23}$ & $17.8 \pm{2.4}$ &$  0.47 \pm{0.09}$  &  $0.06 \pm{0.01}$ &  $ 5.4 \pm{0.8}$ & $1235.4/1283$  \\
60182.9 & $ 0.39 \pm{0.04} $ & $1.95 \pm{0.16}$ & $0.37 \pm{0.03}$ & $0.03 \pm{0.01}$  & $0.39 \pm{0.02}$ & $ 13 \pm{ 4}$  & $1.96 \pm{0.17}$ & $ 6.1 \pm{0.5}$ & $5.42 \pm{0.62}$ & $0.99 \pm{0.15}$ & $0.39 \pm{0.09}$ & $1.36 \pm{0.36}$ & $20.2 \pm{3.7}$ &$  0.52 \pm{0.09}$  &  $0.05 \pm{0.01}$ &  $17.2 \pm{1.7}$ & $1220.6/1283$  \\
60183.1 & $ 0.39 \pm{0.06} $ & $1.90 \pm{0.08}$ & $0.40 \pm{0.03}$ & $0.06 \pm{0.01}$  & $0.34 \pm{0.02}$ & $ 78 \pm{ 4}$  & $1.98 \pm{0.13}$ & $ 6.3 \pm{0.4}$ & $5.29 \pm{0.72}$ & $1.26 \pm{0.13}$ & $0.62 \pm{0.08}$ & $1.35 \pm{0.08}$ & $20.6 \pm{0.7}$ &$  0.08 \pm{0.01}$  &  $0.05 \pm{0.02}$ &  $17.4 \pm{1.2}$ & $1004.8/1313$  \\
60183.8 & $ 0.39 \pm{0.08} $ & $1.76 \pm{0.13}$ & $0.36 \pm{0.02}$ & $0.18 \pm{0.01}$  & $0.28 \pm{0.02}$ & $230 \pm{21}$  & $1.96 \pm{0.15}$ & $ 5.2 \pm{0.6}$ & $5.48 \pm{1.21}$ & $1.05 \pm{0.17}$ & $0.45 \pm{0.06}$ & $1.47 \pm{0.11}$ & $22.2 \pm{1.8}$ &$  0.01 \pm{0.01}$  &  $0.05 \pm{0.02}$ &  $23.9 \pm{2.1}$ & $1149.6/1313$  \\
60184.6 & $ 0.35 \pm{0.08} $ & $1.77 \pm{0.18}$ & $0.35 \pm{0.03}$ & $0.18 \pm{0.02}$  & $0.29 \pm{0.02}$ & $290 \pm{26}$  & $1.99 \pm{0.11}$ & $ 5.7 \pm{0.4}$ & $5.45 \pm{1.17}$ & $1.06 \pm{0.17}$ & $0.51 \pm{0.04}$ & $1.52 \pm{0.12}$ & $22.1 \pm{1.9}$ &$  0.01 \pm{0.01}$  &  $0.05 \pm{0.02}$ &  $27.1 \pm{2.1}$ & $1201.5/1303$  \\
60185.4 & $ 0.35 \pm{0.08} $ & $1.80 \pm{0.08}$ & $0.37 \pm{0.01}$ & $0.18 \pm{0.01}$  & $0.31 \pm{0.04}$ & $210 \pm{14}$  & $2.00 \pm{0.16}$ & $ 6.3 \pm{0.4}$ & $5.34 \pm{1.08}$ & $0.93 \pm{0.17}$ & $0.43 \pm{0.06}$ & $1.53 \pm{0.12}$ & $21.6 \pm{1.9}$ &$  0.01 \pm{0.01}$  &  $0.05 \pm{0.02}$ &  $28.2 \pm{2.1}$ & $1111.1/1285$  \\
60186.2 & $ 0.35 \pm{0.08} $ & $1.76 \pm{0.08}$ & $0.38 \pm{0.04}$ & $0.18 \pm{0.02}$  & $0.33 \pm{0.02}$ & $120 \pm{19}$  & $2.11 \pm{0.15}$ & $ 9.7 \pm{0.5}$ & $5.31 \pm{0.44}$ & $1.02 \pm{0.26}$ & $0.53 \pm{0.07}$ & $1.58 \pm{0.20}$ & $22.0 \pm{2.8}$ &$  0.01 \pm{0.01}$  &  $0.05 \pm{0.01}$ &  $29.1 \pm{2.5}$ & $1122.9/1285$  \\
60187.1 & $ 0.33 \pm{0.06} $ & $1.95 \pm{0.11}$ & $0.40 \pm{0.04}$ & $0.13 \pm{0.01}$  & $0.32 \pm{0.02}$ & $120 \pm{13}$  & $2.10 \pm{0.15}$ & $ 9.5 \pm{0.6}$ & $5.35 \pm{0.21}$ & $0.95 \pm{0.34}$ & $0.45 \pm{0.09}$ & $1.58 \pm{0.36}$ & $21.8 \pm{5.1}$ &$  0.01 \pm{0.01}$  &  $0.05 \pm{0.01}$ &  $28.7 \pm{2.7}$ & $ 996.5/1285$  \\
60187.2 & $ 0.35 \pm{0.08} $ & $1.80 \pm{0.08}$ & $0.39 \pm{0.02}$ & $0.18 \pm{0.02}$  & $0.34 \pm{0.02}$ & $110 \pm{15}$  & $2.13 \pm{0.11}$ & $10.8 \pm{0.5}$ & $5.29 \pm{0.14}$ & $0.98 \pm{0.24}$ & $0.49 \pm{0.07}$ & $1.61 \pm{0.23}$ & $22.6 \pm{3.3}$ &$  0.01 \pm{0.01}$  &  $0.05 \pm{0.01}$ &  $28.7 \pm{2.4}$ & $1078.2/1285$  \\
60188.0 & $ 0.32 \pm{0.04} $ & $1.82 \pm{0.06}$ & $0.40 \pm{0.04}$ & $0.11 \pm{0.01}$  & $0.36 \pm{0.02}$ & $ 41 \pm{ 4}$  & $2.17 \pm{0.12}$ & $15.3 \pm{0.7}$ & $5.31 \pm{0.23}$ & $0.91 \pm{0.04}$ & $0.39 \pm{0.03}$ & $1.63 \pm{0.53}$ & $20.9 \pm{3.3}$ &$  0.50 \pm{0.08}$  &  $0.09 \pm{0.02}$ &  $27.8 \pm{3.3}$ & $1057.8/1313$  \\
60189.0 & $ 0.12 \pm{0.03} $ & $1.95 \pm{0.17}$ & $0.24 \pm{0.02}$ & $0.04 \pm{0.01}$  & $0.37 \pm{0.03}$ & $ 24 \pm{ 4}$  & $2.18 \pm{0.15}$ & $13.5 \pm{0.6}$ & $5.44 \pm{0.13}$ & $1.01 \pm{0.04}$ & $0.43 \pm{0.07}$ & $1.61 \pm{0.53}$ & $22.1 \pm{6.8}$ &$  0.50 \pm{0.01}$  &  $0.05 \pm{0.01}$ &  $27.8 \pm{3.1}$ & $1016.9/1313$  \\
60189.1 & $ 0.12 \pm{0.03} $ & $1.71 \pm{0.05}$ & $0.40 \pm{0.04}$ & $0.11 \pm{0.01}$  & $0.34 \pm{0.05}$ & $ 38 \pm{ 5}$  & $2.14 \pm{0.11}$ & $10.4 \pm{0.5}$ & $5.51 \pm{0.21}$ & $1.01 \pm{0.04}$ & $0.44 \pm{0.06}$ & $1.69 \pm{0.05}$ & $24.4 \pm{0.8}$ &$  0.01 \pm{0.01}$  &  $0.05 \pm{0.01}$ &  $32.1 \pm{2.5}$ & $1466.1/1431$  \\
60190.1 & $ 0.12 \pm{0.04} $ & $1.95 \pm{0.13}$ & $0.24 \pm{0.02}$ & $0.04 \pm{0.01}$  & $0.35 \pm{0.03}$ & $ 37 \pm{ 6}$  & $2.18 \pm{0.16}$ & $11.9 \pm{0.6}$ & $5.54 \pm{0.30}$ & $1.01 \pm{0.08}$ & $0.37 \pm{0.11}$ & $1.74 \pm{0.07}$ & $24.8 \pm{0.1}$ &$  0.17 \pm{0.02}$  &  $0.10 \pm{0.01}$ &  $33.2 \pm{3.1}$ & $1271.4/1431$  \\
60191.0 & $ 0.26 \pm{0.04} $ & $1.95 \pm{0.19}$ & $0.22 \pm{0.03}$ & $0.03 \pm{0.01}$  & $0.36 \pm{0.02}$ & $ 27 \pm{ 3}$  & $2.03 \pm{0.15}$ & $ 6.6 \pm{0.5}$ & $5.47 \pm{0.40}$ & $0.89 \pm{0.04}$ & $0.26 \pm{0.03}$ & $1.66 \pm{0.07}$ & $23.5 \pm{0.1}$ &$  0.22 \pm{0.03}$  &  $0.10 \pm{0.02}$ &  $30.6 \pm{2.9}$ & $1124.7/1431$  \\
60191.1 & $ 0.31 \pm{0.03} $ &        -         &        -         &         -         & $0.41 \pm{0.03}$ & $ 10 \pm{ 1}$  & $2.10 \pm{0.12}$ & $ 8.4 \pm{1.2}$ & $5.51 \pm{0.43}$ & $1.01 \pm{0.13}$ & $0.35 \pm{0.05}$ & $1.68 \pm{0.08}$ & $25.1 \pm{0.9}$ &$  0.01 \pm{0.01}$  &  $0.05 \pm{0.02}$ &  $30.4 \pm{1.5}$ & $1250.1/1434$  \\
60192.2 & $ 0.34 \pm{0.06} $ &        -         &        -         &         -         & $0.43 \pm{0.04}$ & $ 10 \pm{ 1}$  & $2.04 \pm{0.11}$ & $ 6.2 \pm{1.0}$ & $5.52 \pm{0.43}$ & $1.01 \pm{0.11}$ & $0.35 \pm{0.04}$ & $1.64 \pm{0.12}$ & $25.4 \pm{1.1}$ &$  0.01 \pm{0.01}$  &  $0.05 \pm{0.02}$ &  $27.8 \pm{2.9}$ & $1232.7/1434$  \\
60193.4 & $ 0.25 \pm{0.04} $ &        -         &        -         &         -         & $0.44 \pm{0.05}$ & $  6 \pm{ 3}$  & $2.25 \pm{0.09}$ & $16.8 \pm{0.8}$ & $5.47 \pm{0.42}$ & $1.01 \pm{0.11}$ & $0.41 \pm{0.07}$ & $1.65 \pm{0.07}$ & $24.9 \pm{0.8}$ &$  0.01 \pm{0.01}$  &  $0.10 \pm{0.01}$ &  $23.7 \pm{1.8}$ & $1496.2/1434$  \\
60194.1 & $ 0.35 \pm{0.06} $ &        -         &        -         &         -         & $0.42 \pm{0.03}$ & $ 11 \pm{ 6}$  & $2.08 \pm{0.14}$ & $ 7.5 \pm{1.2}$ & $5.46 \pm{0.42}$ & $0.69 \pm{0.11}$ & $0.19 \pm{0.03}$ & $1.66 \pm{0.07}$ & $25.6 \pm{0.8}$ &$  0.01 \pm{0.01}$  &  $0.05 \pm{0.01}$ &  $26.9 \pm{1.1}$ & $1177.1/1434$  \\
60194.5 & $ 0.12 \pm{0.02} $ &        -         &        -         &         -         & $0.45 \pm{0.04}$ & $  3 \pm{ 1}$  & $2.20 \pm{0.16}$ & $10.9 \pm{0.4}$ & $5.57 \pm{0.48}$ & $1.09 \pm{0.28}$ & $0.41 \pm{0.08}$ & $1.84 \pm{0.14}$ & $29.5 \pm{2.1}$ &$  0.52 \pm{0.03}$  &  $0.10 \pm{0.01}$ &  $32.1 \pm{1.1}$ & $1697.1/1428$  \\
60195.0 & $ 0.35 \pm{0.08} $ &        -         &        -         &         -         & $0.42 \pm{0.04}$ & $ 10 \pm{ 1}$  & $2.02 \pm{0.11}$ & $ 5.5 \pm{0.9}$ & $5.52 \pm{0.47}$ & $0.71 \pm{0.19}$ & $0.15 \pm{0.07}$ & $1.68 \pm{0.22}$ & $26.4 \pm{2.2}$ &$  0.01 \pm{0.01}$  &  $0.05 \pm{0.01}$ &  $27.9 \pm{1.1}$ & $1048.9/1432$  \\
60195.2 & $ 0.35 \pm{0.08} $ &        -         &        -         &         -         & $0.42 \pm{0.03}$ & $  9 \pm{ 1}$  & $2.01 \pm{0.13}$ & $ 5.5 \pm{1.0}$ & $5.51 \pm{0.42}$ & $0.73 \pm{0.11}$ & $0.17 \pm{0.05}$ & $1.68 \pm{0.21}$ & $26.5 \pm{2.1}$ &$  0.01 \pm{0.01}$  &  $0.05 \pm{0.01}$ &  $27.9 \pm{1.1}$ & $1070.8/1432$  \\
60195.6 & $ 0.12 \pm{0.03} $ &        -         &        -         &         -         & $0.38 \pm{0.03}$ & $  6 \pm{ 1}$  & $2.16 \pm{0.36}$ & $ 8.9 \pm{1.1}$ & $5.59 \pm{0.43}$ & $1.01 \pm{0.13}$ & $0.41 \pm{0.08}$ & $1.81 \pm{0.08}$ & $30.5 \pm{0.7}$ &$  0.42 \pm{0.04}$  &  $0.10 \pm{0.01}$ &  $29.1 \pm{2.2}$ & $1515.8/1430$  \\
60196.0 & $ 0.35 \pm{0.08} $ &        -         &        -         &         -         & $0.44 \pm{0.05}$ & $  8 \pm{ 1}$  & $2.05 \pm{0.15}$ & $ 6.2 \pm{1.0}$ & $5.52 \pm{0.47}$ & $0.74 \pm{0.17}$ & $0.15 \pm{0.02}$ & $1.72 \pm{0.10}$ & $27.8 \pm{1.6}$ &$  0.01 \pm{0.01}$  &  $0.05 \pm{0.01}$ &  $28.6 \pm{3.3}$ & $1048.1/1434$  \\
60196.2 & $ 0.35 \pm{0.08} $ &        -         &        -         &         -         & $0.43 \pm{0.04}$ & $  8 \pm{ 1}$  & $2.06 \pm{0.16}$ & $ 6.6 \pm{1.1}$ & $5.55 \pm{0.47}$ & $0.66 \pm{0.17}$ & $0.13 \pm{0.04}$ & $1.69 \pm{0.10}$ & $26.7 \pm{1.7}$ &$  0.06 \pm{0.01}$  &  $0.05 \pm{0.01}$ &  $27.4 \pm{2.2}$ & $1036.3/1434$  \\
60196.6 & $ 0.12 \pm{0.03} $ &        -         &        -         &         -         & $0.35 \pm{0.03}$ & $ 12 \pm{ 2}$  & $2.05 \pm{0.18}$ & $ 5.5 \pm{0.8}$ & $5.61 \pm{0.47}$ & $1.01 \pm{0.17}$ & $0.39 \pm{0.04}$ & $1.76 \pm{0.11}$ & $29.8 \pm{1.8}$ &$  0.44 \pm{0.03}$  &  $0.10 \pm{0.01}$ &  $27.6 \pm{1.9}$ & $1517.3/1434$  \\
60197.0 & $ 0.34 \pm{0.08} $ &        -         &        -         &         -         & $0.43 \pm{0.03}$ & $  8 \pm{ 1}$  & $1.99 \pm{0.10}$ & $ 4.4 \pm{0.7}$ & $5.57 \pm{0.40}$ & $0.71 \pm{0.24}$ & $0.14 \pm{0.04}$ & $1.69 \pm{0.17}$ & $28.6 \pm{2.5}$ &$  0.14 \pm{0.01}$  &  $0.05 \pm{0.01}$ &  $27.1 \pm{3.1}$ & $1009.5/1434$  \\
60197.1 & $ 0.35 \pm{0.05} $ &        -         &        -         &         -         & $0.43 \pm{0.04}$ & $  8 \pm{ 1}$  & $2.06 \pm{0.13}$ & $ 6.8 \pm{1.1}$ & $5.44 \pm{0.20}$ & $0.75 \pm{0.32}$ & $0.16 \pm{0.01}$ & $1.73 \pm{0.32}$ & $27.2 \pm{4.6}$ &$  0.31 \pm{0.01}$  &  $0.05 \pm{0.01}$ &  $27.5 \pm{1.9}$ & $1108.2/1434$  \\
60197.6 & $ 0.12 \pm{0.03} $ &        -         &        -         &         -         & $0.44 \pm{0.02}$ & $  3 \pm{ 7}$  & $2.20 \pm{0.15}$ & $10.5 \pm{1.0}$ & $5.61 \pm{0.14}$ & $1.05 \pm{0.19}$ & $0.35 \pm{0.12}$ & $1.83 \pm{0.21}$ & $31.7 \pm{3.1}$ &$  0.50 \pm{0.02}$  &  $0.10 \pm{0.01}$ &  $28.2 \pm{0.8}$ & $1455.9/1434$  \\
60198.0 & $ 0.35 \pm{0.04} $ &        -         &        -         &         -         & $0.43 \pm{0.03}$ & $  9 \pm{ 1}$  & $2.14 \pm{0.13}$ & $ 8.1 \pm{0.7}$ & $5.52 \pm{0.21}$ & $0.55 \pm{0.04}$ & $0.11 \pm{0.03}$ & $1.89 \pm{0.44}$ & $31.9 \pm{2.9}$ &$  0.50 \pm{0.02}$  &  $0.09 \pm{0.01}$ &  $35.7 \pm{3.4}$ & $1087.4/1434$  \\
 \hline
 \end{tabular}}
 \noindent{
 \leftline{Column 1 represents the MJD of those respective Exposure IDs for which we have performed spectral analysis.}
 \leftline{Column 2 gives the values of hydrogen column densities ($N_H$) of those analyzed exposures.}
 \leftline{Columns 3, 4, \& 5 give the values of the parameters of the `gabs' model.}
 \leftline{Columns 6 \& 7 give the model fitted values of the parameters from the `diskbb' model. The Norm is a value of the ratio of the inner disk radius ($R_{in}$) in $km$, source distance ($D_{10}$) in 10 kpc, and cosine of the source inclination angle.}
 \leftline{Columns 8 \& 9 give the model fitted values of the parameters from the `power-law' model.}
 \leftline{Columns 10, 11, \& 12 give the model fitted values of the parameters from the `Gaussian' model.}
 \leftline{Columns 13--17 give the model fitted values of the parameters from the `pexrav' model. reflection scaling factor ($R_{ref}$, $= 0$ for no reflected component, $<$0 reflection component only).}
 \leftline{Column 18 gives $\chi^2/DOF$ values of the respective spectral fits.}
 \leftline{The errors are estimated with 90\% confidence interval, that corresponds to $1.645 \sigma$ in XSPEC.}
 }
\end{sidewaystable}

\end{document}